\documentclass[ACS,STIX1COL]{WileyNJD-v2}

\articletype{Article}%

\received{\today}
\revised{}
\accepted{}

\usepackage{color}
\definecolor{rev}{rgb}{0,0,0} % set {0,0,0} for black ink

\begin{document}

%\title{This is the sample article title\protect\thanks{This is an example for title footnote.}}

\title{Hybrid analysis and modeling, eclecticism, and multifidelity computing toward digital twin revolution}

\author[1]{Omer San*}

\author[2,4]{Adil Rasheed}

\author[3,4]{Trond Kvamsdal}

%\authormark{OMER SAN \textsc{et al}}

\address[1]{\orgdiv{School of Mechanical \& Aerospace Engineering}, \orgname{Oklahoma State University}, \orgaddress{\state{Stillwater, OK 74078}, \country{USA}}}

\address[2]{\orgdiv{Department of Engineering Cybernetics}, \orgname{Norwegian University of Science and Technology}, \orgaddress{\state{7465 Trondheim}, \country{Norway}}}

\address[3]{\orgdiv{Department of Mathematical Sciences}, \orgname{Norwegian University of Science and Technology}, \orgaddress{\state{7491 Trondheim}, \country{Norway}}}

\address[4]{\orgdiv{Department of Mathematics and Cybernetics}, \orgname{SINTEF Digital}, \orgaddress{\state{7034 Trondheim}, \country{Norway}}}

\corres{*Omer San, School of Mechanical \& Aerospace Engineering, Oklahoma State University, Stillwater, OK 74078, USA. \email{osan@okstate.edu}}

%\presentaddress{This is sample for present address text this is sample for present address text}

\abstract{Most modeling approaches lie in either of the two categories: physics-based or data-driven. Recently, a third approach which is a combination of these deterministic and statistical models is emerging for scientific applications. To leverage these developments, our aim in this perspective paper is centered around exploring numerous principle concepts to address the challenges of (i) trustworthiness and generalizability in developing data-driven models to shed light on understanding the fundamental trade-offs in their accuracy and efficiency, and (ii) seamless integration of interface learning and multifidelity coupling approaches that transfer and represent information between different entities, particularly when different scales are governed by different physics, each operating on a different level of abstraction. Addressing these challenges could enable the revolution of digital twin technologies for scientific and engineering applications.}


\maketitle

%\footnotetext{\textbf{Abbreviations:} HAM, hybrid analysis and modeling; SML, scientific machine learning; DRL, deep reinforcement learning}

\section{Introduction}\label{sec:intro}
According to Peter Shaver \cite{shaver2018rise} there have been two monumental steps in the rise of science: 1) The \textit{qualitative and descriptive} worldview of Aristotle, 2) The \textit{quantitative and predictive} worldview of Newton. Newton's laws of motion is the foundation for physics-based models in engineering and earth science that we now may complement with artificial intelligence (AI) and machine learning (ML) techniques through interaction with the physical reality in question by means of real-time sensor data. We consider the combined use of physics-based and data-driven models, herein denoted \emph{hybrid analysis and modeling} (HAM), as a third monumental step of science characterised by being \textit{interactive and evolving}. \emph{Interactive} in the sense that the combined model is based both on a priori human perception of the physical reality as well as on real-time experimental (through sensors) observations. \textit{Evolving} in the sense of changing with time according to more available observational data utilised by the data-driven methods as well as changes of the involved objects (e.g. degradation) and processes. A key ingredient in our perspective is to develop HAM tools for scientific and engineering applications and make these techniques available for industry to be used in design and operation by means of developing highly capable \emph{digital twins} (DTs). We postulate and foresee that the HAM approach will be of disruptive nature for \emph{computational science and engineering} (CSE) and initiate an exponential growth of scientific/engineering knowledge and innovations, and that DTs will be the natural means to facilitate the take up of the new knowledge and innovations into design and operation by the industry.

Although a healthy criticism about \emph{data-driven modeling} goes around in scientific communications, ML has started to become a new tradition in many scientific disciplines. If we are not able to understand, let's learn it! However, if not well grounded, such data-driven learning might yield a \emph{perpetual engine} that can be viewed as a sort of \emph{fishing expedition}, especially for modeling large scale systems. Moreover, if there is already a \emph{white-box} principled procedure available to model a certain process, it is arguably safer that we should avoid using a black-box approach that is blindly reconstructed from data. On the other hand, the trade-off between accuracy and efficiency could be a game changer for numerous applications. Yet, for example, it might not be that far from today to use applications in our smart phones to contribute value of crowd sourced data in improving operational weather forecasts. More importantly, in some cases, we might lack having a principled approach for an underlying process, hence data becomes essential to build a model. Crucially, we highlight that the data-driven models need to possess important characteristics (e.g., symmetry, conservation, stability, and scalability properties) in order to revamp the existing predictive tools. To address some of these challenges, here we look into \emph{eclecticism} as a key computing philosophy, and explore an emerging HAM framework that combines the generalizability and interpretability of physics-based models with the versatility of data-driven modeling. This can be viewed as a divide-compute-merge-conquer approach, for which the \emph{compute} part highly benefits from the advances in computational mathematics, and the \emph{merge} part utilizes the evolutionary shift in modern computational infrastructures to effectively account for representation losses and closures in multi-x workflows. Herein the \emph{multi-x} refers to the systems where the underlying models involve multiple characteristic scales, multiple spatiotemporal domains, multiple physical laws, and even multiple disciplines. \textcolor{rev}{\emph{Divide} part might consist clustering, spatial or spatio-temporal partitioning, and we term \emph{conquer} in referring to the acceleration of the overall process or improving its system level predictive performance. For example, it is very common to use a coarse-grid solver to resolve the atmospheric boundary layer flow and employ the fine-grid solver to resolve the flow field around wind turbines in the wind farm. For such problems, it is very important to exchange the information between two solvers that leads to physically consistent boundary conditions, and statistical tools like deep learning can be exploited for coupling different solvers and mathematical descriptions.}

In a nutshell, our prospective survey aims at exploring imminent algorithmic needs in \emph{predictive engineering} for multi-x systems; complex natural or engineered systems comprising multiple physical or computational heterogeneous entities. Such predictive engines serve as the inner-workhorse for outer-workflow loops such as optimal design, control, and discovery. While immense advances in computational mathematics and scientific computing have come to fruition, these methods often suffer a curse of dimensionality limiting turnaround. The field of \emph{multifidelity computing} therefore aims to address this computational challenge by exploiting the relationship between high-fidelity and low-fidelity models. Here, we focus on physical systems where the computational domain consists of a set of nested hierarchical models or solvers. Examples range from systems within the Earth’s core, its shallow subsurface and atmosphere, and beyond. A grand challenge in modeling and computing such complex multi-x systems is the difficulty of determining proper interface boundary conditions, and consistent restriction and prolongation operators among solvers. We often come to a point in modeling where we ask the questions about the legitimate ways to exploit the use of statistical inference algorithms to fuse two heterogeneous computational entities having an interface, which often consist of nested models, multiple physics, and parametric and spatiotemporal heterogeneity. This survey intends primarily to share our perspectives on some of these potential questions.  

\textit{Eclecticism}, according to the Cambridge English Dictionary, refers to a conceptual philosophical approach that does not hold rigidly to a single paradigm or set of assumptions, but instead draws upon multiple theories, formulations, or ideas to gain complementary insights into a subject, or applies different theories in particular cases. In computational sciences, eclecticism may mean combining whatever seems the best or most useful to improve predictions against limited computational resources. This can be effectively achieved using a \emph{domain decomposition} approach \cite{bjorstad1986iterative,chan1989domain,Farhat1991amo, chan1991domain,gropp1992domain,chan1994domain,li2006feti,toselli2006domain,gander2008schwarz,pavarino2012recent,tu2015feti,dolean2015introduction,pedneault2017schur,houzeaux2017domain,tang2020review}, a powerful concept often used to solve partial differential equations (PDEs) with complex geometries. Here, we can interpret the term \emph{domain decomposition} in a broad sense such that it also includes the simultaneous advances in the multiscale and multiphysics problems \cite{Hughes1995mpg, Hughes1998tvm, weinan2007heterogeneous,pavliotis2008multiscale,fish2010multiscale,matouvs2017review,geers2017homogenization,soane2019multigrid,you2020asymptotically,wheeler1998physical,cao2014parallel,alber2019integrating}. Many approaches in the field of domain decomposition are well-suited for heterogeneous problems, which have dissimilar governing equations or models (a.k.a. abstractions) in different regions of interest. 

In terms of classification, two main coupling approaches are devised to merge these abstractions \cite{keyes2013multiphysics}: (i) the monolithic or intrusive framework \cite{gosselet2004monolithic,barker2010scalable,gee2011truly,langer2016recent,sauer2018monolithic,heinlein2019monolithic} that generates a unified system for the coupled problem, and (ii) the partitioned or non-intrusive framework \cite{Felippa1980sta, quarteroni1999domain,Felippa2001pao, tezduyar2007modelling,kuttler2008fixed,ha2017comparative,quarteroni2017domain} in which we solve multiple abstractions separately, and exchange interface boundary conditions between them (a.k.a. transmission conditions) using sequential or parallel Schwarz iterations \cite{lions1988schwarz,lions1989schwarz,lions1990schwarz}. In practice, the latter is often preferred due to the flexibility of using pre-existing solvers for each abstraction, which is also the approach we will pursue in accomplishing \emph{scalable eclecticism}. Two conflicts are in order. First, the iterative nature of this approach poses significant challenges on the scalability of the ultimate joint solver, even though an extreme scalability may be already presented in the individual abstractions. Second, although optimal transmission conditions can be derived for linear or simplified model systems \cite{tan1994generalized,engquist1998absorbing,japhet1998optimized,discacciati2013interface,gander2015optimized,blayo2016towards,gander2019heterogeneous}, it is often unclear how good these transmission conditions are for realistic applications, especially when we consider their accuracy, stability, and convergence rate characteristics. However, these conflicts offer new challenges as well as opportunities for researchers working on the next-generation statistical inference approaches for scientific applications. We arguably hypothesize that it might be viable to introduce statistical inference tools into multi-x workflows to develop robust predictive engines that require no iteration between such abstractions. \textcolor{rev}{Within the context of domain decomposition, we refer readers to recent discussions on exploratory studies on ML to couple numerical solutions of partial differential equations \cite{tang2021exploratory,li2019d3m}. Moreover, neural networks have been utilized to predict the geometric location interfaces in developing adaptive domain decomposition methods \cite{heinlein2019machine}. As highlighted in our recent works \cite{ahmed2020interface,ahmed2021multifidelity}, ML tools can be effectively utilized in various forms of interfacial error correction to form the building blocks of an integrated approach among mixed fidelity descriptions toward predictive DT technologies. 
}

From modeling perspective, overall, two distinct approaches are in vogue in the scientific community today: physics-based modeling and data-driven modeling. Both these approaches have their own strengths and weaknesses and hence in the present paper we argue for a need to combine these two modeling approaches in ways that \textcolor{rev}{benefit from the best of both approaches.} To this end, we aim to highlight tentative ways of combining them. We also stress that the coupling of these complementary approaches in an eclectic way also poses novel computational, mathematical and phenomenological issues that should be explored.  So far, the computational science and engineering community has been driven mostly by a \emph{physics-based modeling} approach. This approach consists of observing physical phenomena, developing approximate mathematical models \textcolor{rev}{from first principles,} and ultimately solving them. We often observe only a part of the physics, and model even a smaller part of underlying phenomena. To make the simulations computationally tractable, more assumptions are made leading to further loss of physics. In any physics-based approach, therefore, we take into account only a fraction of all the physical processes, and that partly explains the mismatch between predictions and observations. On the other hand, we have a purely \emph{data-driven modeling} approach that thrives on the availability of data. It is expected that data is a manifestation of all the physics driving a particular process and hence even without any knowledge of the governing physics, one can model their effects. This approach, which was challenged until recently, is now gaining attention primarily because of the access to unprecedented amounts of data, open source cutting-edge ML and data analytics libraries, and computational resources. Together with their modularity and simplicity, data-driven models also offer a unique advantage in multidisciplinary collaborative environments and provide richer platforms securing intangible assets and intellectual property rights \cite{ahmed2019memory,pawar2019deep}. As we highlight in our technology watch survey \cite{rasheed2020digital}, such an interconnected HAM paradigm constitutes a key enabler for the emerging DT revolution. In this perspective letter, we aim to provide an overview of HAM strategies relevant to scientific and engineering applications. The topic spans a wide spectrum, and there are a great number of review articles on relevant discussions, methodologies and applications \cite{kashinath2021physics,bauer2021digital,geer2021learning,brunton2020machine,taira2020modal,frank2020machine,zhu2020river,tahmasebi2020machine,willard2020integrating,mendoncca2019model,yu2019non,brenner2019perspective,yondo2019review,taira2017modal,rowley2017model,benner2015survey,brunton2015closed,asher2015review,lassila2014model,machairas2014algorithms,mezic2013analysis,mignolet2013review,bazaz2012review,razavi2012review,theodoropoulos2011optimisation,chinesta2011short,wagner2010model,massarotti2010reduced,kleijnen2009kriging,pinnau2008model,lucia2004reduced,ong2003evolutionary,freund2003model,bai2002krylov,chatterjee2000introduction,berkooz1993proper,bonvin1982unified,elrazaz1981review}. Therefore, it is not our intention to give a complete biography, but rather present somehow our \emph{subjective} perspectives with an emphasis on the emerging methodologies and enabling technologies from modeling perspectives. In particular, there has been recent interest in \emph{big data}, AI, and ML applications in numerous disciplines, and therefore our survey is intended to encourage cross-disciplinary efforts between many stakeholders including practitioners, physicists, mathematicians and data scientists.

We emphasize that advanced ML models like deep learning networks are powerful tools for finding patterns in complicated datasets. However, as these \emph{universal approximators}\cite{hornik1989multilayerua,hagan1994training,hagan1997neural,vidal2017mathematics} become complex, the number of trainable parameters (weights) quickly explodes, adversely affecting their interpretability and hence their trustworthiness.  Using these models in combination with other traditional models compromises the trustworthiness of the overall system. Some of the active areas of research in this context are:  cost function modification to accommodate the model Jacobian \cite{lee1997hybrid}, grow-when-required network \cite{marsland2002self}, physics-informed neural networks (PINNs) \cite{raissi2019physics,zhu2019physics,pan2020physics}, embedding \emph{hard} physical constraints in a neural network\cite{mohan2020embedding}, leveraging uncertainty information \cite{leibig2017leveraging}, developing visualization tool for the network analysis \cite{sacha2017you}, and nonparametric modeling approaches for bridging data science and dynamical systems \cite{berry2020bridging}. While these techniques have been emerging in both scientific computing and ML fields, they offer many opportunities to fuse topics in numerical linear algebra and theoretical computer science \cite{golub2006bridging} towards improving the explainability of these models and implementing built-in sanity checks on the overall system where these models are employed. With this in mind, we present our perspectives to provide a glimpse into this exciting field, which came to known as \emph{HAM}.

\textcolor{rev}{
This paper is organized as follows. In Section~\ref{sec:ham}, we describe how HAM approaches can tackle important issues in physics-based or data-driven models, for instance, how the generalization properties of data-driven models can be improved by using domain knowledge. Sections~\ref{sec:pgml} and \ref{sec:il} are devoted to two promising HAM frameworks, physics-guided ML and interface learning, respectively. Sections~\ref{sec:rom} and \ref{sec:big} illustrate the application of HAM approaches in reduced order modeling and big data cybernetics, and Section~\ref{sec:dt} describes the idea of DTs and different capability stages of DTs. The final section then gives concluding remarks and an outlook with remarks on open topics in the field of HAM.
}

\section{Hybrid analysis and modeling}\label{sec:ham}

\emph{Quo vadis} in modeling: physics-based or data-driven? One camp develops and improves first principle models that provide the best generalizability and trustworthiness characteristics. In the other camp, researchers try to explain phenomena from archival data using statistical approaches. 
With the abundant supply of big data, open-source cutting edge and easy-to-use ML libraries, data-driven modeling has become very popular. 
Compared to the physics-based modeling approach, these models thrive on the assumption that data is a manifestation of both known and unknown physics and hence when trained with an ample amount of data, the data-driven models might learn the full physics on their own. This approach, involving in particular deep learning, has started achieving human-level performance in several tasks. Notable among these are image classification \cite{ciresan2012multi,litjens2017a,szegedy2017inception,li2019deep}, dimensionality reduction \cite{Hinton504}, medical treatment \cite{shen2017deep,xu2019deep,LIU20191,antil2020bilevel}, smart agriculture \cite{BU2019500}, physical sciences \cite{brunton2020machine,reichstein2019deep,schmidt2019recent} and beyond. Some of the advantages of these models are online learning capability, computational efficiency for inference, accuracy even for very challenging problems as far as the training, validation and test data are prepared properly. 
However, due to their data-hungry and black-box nature, poor generalizability, inherent bias and lack of robust theory for the analysis of model stability, their acceptability in multi-x systems is fairly limited. In fact, the numerous vulnerabilities of deep neural networks have been exposed beyond doubt in several recent works \cite{akhtar2018threat,yuan2019adversarial,hao2020adversarial}. To address such challenges, we explore a new HAM approach based on a synergistic combination of deterministic and statistical model components.

\begin{figure}[ht]
\centering
\includegraphics[trim= 0 0 0 0, clip, width=0.95\textwidth]{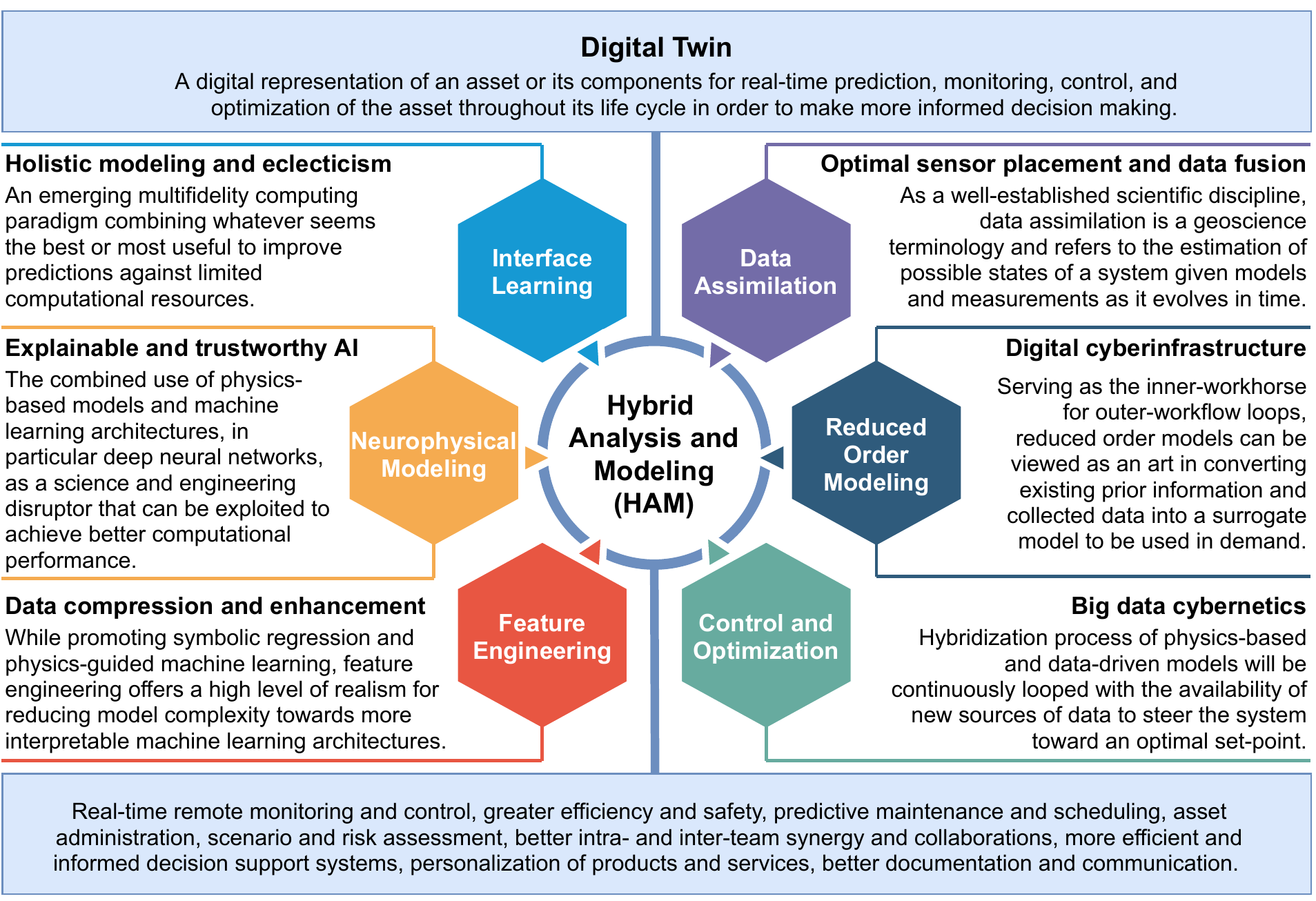}
\caption{Overview of the \emph{HAM} paradigm as a science and engineering disruptor while balancing technology push and domain-specific application pull. \textcolor{rev}{As a core methodological facilitator in combining first principles with data-driven approaches, the HAM framework benefits from the advances in both modeling disciplines towards building more explainable and trustworthy predictive tools. Since DTs often comprise heterogeneous computational entities to maximize the utilization of available resources, the interface learning paradigm allows seamless coupling of these individuals components while neurophysical modeling approach aims at elevating the reliability and computational efficiency of such entities by combining physics-based and data-driven methodologies. For the latter, feature engineering improves their interpretability by minimizing the number of trainable parameters and reducing the size of search space. Data assimilation can also be viewed as one of the well-established HAM algorithms that fuse models and data to cure the deficiencies in both, which can be further implemented for prediction and control purposes.  Nonetheless, their inner workhorses are often computationally expensive for realistic applications. Fortunately, reduced order modeling provides light surrogates to mitigate this burden, where again HAM plays an important role in improving the accuracy, stability, and trustworthiness of these proxy models. The integration and synergy of all these pave the way towards the realization of next-generation DT technologies in the first place.
}}
\label{fig:ham}
\end{figure}

We term HAM as a modeling approach that combines the interpretability, robust foundation and understanding of a physics-based approach with the accuracy, efficiency, and automatic pattern-identification capabilities of advanced data-driven ML and artificial intelligence algorithms. Our overall hypothesis for developing hybrid methodologies is to form a solid foundation that balances technology push and domain-specific application pull. In particular, we consider the combined use of physics-based and data-driven models in a broad sense, hereafter denoted as HAM, as a science and engineering disruptor that can be discerned in Figure~\ref{fig:ham}. \textcolor{rev}{In this figure, we display the different ideas that can be considered the pillars of the HAM paradigm. The interface learning framework allows us to take advantage of models of different levels of fidelity to optimize the available computational resources. The neurophysical modeling framework aims at bringing the best of physics-based and data-driven approaches to deliver computationally economical and generalizable robust models. The reduced order modeling framework can facilitate the construction of efficient surrogate models that can serve as the inner-workhorse of outer loop applications like model predictive control. The data assimilation framework lets us continuously update the models built from the HAM paradigm by exploiting online measurements to reach the optimal set point. Feature engineering, which is a process of using domain knowledge to extract features from data, is critical to all these frameworks that guide the overall design of hybrid models with the objective of bringing trustworthiness in hybrid models. Finally, all these ideas can be interlinked or utilized in isolation to bring realism to a DT of any physical system.}      

%Figure~\ref{fig:ham} shows a non-inclusive list of various directions where HAM can be pursued. 

To illustrate the HAM concept at an abstract level, let's consider a generic prognostic equation relevant to general circulation models (GCMs):
\begin{equation} \label{eq:main}
\frac{\partial \boldsymbol u}{\partial t} + \mathcal{F}(\boldsymbol u; \boldsymbol x, t) = \mathcal{S}(\boldsymbol u; \boldsymbol x, t) 
\end{equation}
where $\boldsymbol u$ is a three-dimensional (3D) dependent variable (e.g., density, velocity, temperature, moisture); $\boldsymbol x$ is a 3D independent variable (i.e., latitude, longitude, height); $\mathcal{F}$ is the model's dynamical core (i.e., a semi-discretized function of PDEs of mass, momentum, energy conservation laws); $\mathcal{S}$ is the model's physics (a.k.a. parametrization) that refers to physical and chemical processes (e.g., turbulence, precipitation, long and short wave atmospheric radiation, clouds, constituency transport, and chemical reactions). Although most physical processes have been included in GCMs as their simplified one-dimensional (1D) parameterized versions (in the vertical direction), still computing $\mathcal{S}$ takes about 70\%--90\% time of the total forward model computations \cite{krasnopolsky2006new,krasnopolsky2006complex,collins2001parameterization,belochitski2011tree}.  
%We highlight that most physical processes have been included in GCMs as their simplified 1D parameterized versions (in the vertical direction), still computing $\mathcal{S}$ takes about 70\%--90\% of the total forward model computations \cite{krasnopolsky2006new,krasnopolsky2006complex,belochitski2011tree}, even if, in practice, model dynamics $\mathcal{F}$ is computed at every $\delta t = 20$ minutes, while some parameterizations (e.g., long wave radiation process that requires a solution of integro-differential equations over a large wavenumber \cite{collins2001parameterization}) in $\mathcal{S}$ is only calculated every $\Delta t = 12$ hours. 
Moreover, these simulations generate a huge volume of climate data that poses storage challenges. %, e.g., 260 terabytes of new data every 16 seconds. 
%Moreover, these simulations generate huge volume of climate data that poses storage challenges (e.g., 260 terabytes of new data every 16 seconds).
This motivates us to develop a HAM approach to improve the handling of $\mathcal{S}$ through the \emph{discovery of better parameterizations}, and the \emph{data-driven acceleration} approaches in the calculation of model physics components. Another major modeling challenge that drives us to establish a bridge between physics-based models and data-driven models is the difficulty of transferring source term effects from a local model to the global model (e.g., turbulence flux, momentum flux, and energy flux) or vice versa for which the global model provides boundary conditions to the local one. 

More specifically, we split $\mathcal{S}$ as 
\begin{equation} 
\label{eq:mainSource}
 \mathcal{S}(\boldsymbol u; \boldsymbol x, t) = \mathcal{D}(\boldsymbol u; \boldsymbol x, t) + \mathcal{N}(\boldsymbol u; \boldsymbol x, t)
\end{equation}
%\vspace{-1ex}
%\begin{equation} \label{eq:main}
%\frac{\partial \boldsymbol u}{\partial t} + \mathcal{F}(\boldsymbol u; \boldsymbol x, t) = \mathcal{D}(\boldsymbol u; \boldsymbol x, t) + %\mathcal{N}(\boldsymbol u; \boldsymbol x, t)
%\vspace{-1ex}
%\end{equation}
where $\mathcal{D}$ refers to the deterministic part of the physics that is manageable to compute or already well established, and $\mathcal{N}$ refers to uncertain processes representing model physics, or refers to a crude approximation that can be improved. 
Recent studies by the authors \cite{pawar2020evolve,pawar2020data,rahman2018hybrid, san2018neural,maulik2019sub} and other researchers \cite{pathak2018hybrid,vlachas2018data,wan2018data,watson2019applying,wikner2020combining,hamilton2017hybrid,wu2018physics,rackauckas2020universal,scher2019weather,hsieh1998applying,tang2001coupling,chevallier1998neural,krasnopolsky2002neural,krasnopolsky2005new} show that  feasible learning is possible for $\mathcal{N}$ that improves the overall accuracy of the forward model, in which a computational speed-up of several orders of magnitude has been achieved. By replacing $\mathcal{N}$ with trainable neural networks and symbolically discovered parameterizations, we can create a modular HAM framework that offers opportunities in modeling and computing many complex systems via developing (i) more sophisticated physical parameterizations, (ii) more frequent incorporation of model physics, (iii) allowing the dynamical core to run with higher resolution, (iv) easier ensemble simulations with reduced uncertainties, and (v) alleviating computational complexity of running adjoint (tangent linear) models when integrated within variational data assimilation workflows.       
%Since improving $F$ has been already exploited significantly by computational mathematicians and physicists over the past few decades, most of recent focus has
%Furthermore, the amount of climate data produced by scientists grows every second: one recent climate simulation required the generation of 260 terabytes of data every 16 seconds. Both the huge volume of existing data and the rapid generation of new data pose immense challenges to our storage system. 
%We propose the following key research questions to establish the foundations of our hybrid approach: 
The outlook of the HAM approach can be established by tackling the following key questions:

\medskip
\emph{How to manage and process data in real-time?}
%\hl{Already: Data generation rate > data communication rate, Data generated so far > data storage capacity: Solutions: compressed sensing, optimal sensor placement, data compression}
Real-time control, for example, is necessary for environmental situational awareness for improved decision-making scenarios, such as wind-turbines adaptation \cite{benjamin2016north}. To achieve this, an increasing number of observations should be assimilated instantaneously since the lifespan of many weather phenomena can be less than a few minutes. Using a full order model (FOM) \citep{NORDANGER2015183, NORDANGER2015664, NORDANGER2016324} with the required resolution is generally not compatible with the time-scales of natural phenomena in realistic scenarios. Therefore, a computationally efficient \emph{reduced order model} (ROM) \citep{Almroth1978aco, Noor1980rbt, Moore1981pca, Ammar2007anf, Rozza2008rba, Kalashnikova2010ots, Chinesta2011asr, quarteroni2014reduced, lassila2014model,  Ballani2018acb, stabile2019reduced, fonn2019fast}, capable of duplicating the physical system with acceptable accuracy, could offer a realistic solution to assimilate these short-time observations. By using adaptive finite element FOM\citep{Babuska1978eef, Babuska1978ape, Ladeveze1983eep, Zienkiewicz1987ase, Ainsworth1989aot, Ainsworth1992ape, Zienkiewicz1992tsp1, Zienkiewicz1992tsp2, Johnson1992afe, Paraschivoiu1997apf, Kvamsdal1998eeb, Okstad1999spr, Ainsworth2000ape, Babuska2001tfe, Melbo2003goe, Bangerth2013afe, Johannessen2014iau, Kumar2015sap, Kumar2017spr} (in the offline stage) together with certified ROM\citep{Porsching1985eot, Rozza2008rba, Machiels2001obf, Eftang2011ahc, Eftang2011ats, Haasdonk2011erm, Knezevic2011rba,  Huynh2011hfr, Huynh2012crb} (in the online stage) we may indeed control the numerical discretization error within prescribed tolerances.
However, often the role of ROM is to give quick guidelines on whether or not a decision needs to be taken. This is specifically important given the intrinsic stochastic characteristics of weather systems, since the model should be able to monitor those seemingly-random sequential observations \cite{dee1998data,dee2000data,chinesta2018virtual}. The use of ROMs in real-time simulations becomes a promising approach in several applications \cite{badias2018reduced,aguado2015real,aguado2017simulation,niroomandi2012accounting}. 
Data-driven tools do an excellent job in terms of online time-cost, however, relying only on data and disregarding the known and well-established physics is not a wise step. Hybridization techniques have been demonstrated to give superior performance to either individual components \cite{krasnopolsky2005new,krasnopolsky2006complex,rahman2018hybrid,pathak2018hybrid,reichstein2019deep,ahmed2021nudged}. 
Therefore, a DT based on HAM methodologies can be built to provide nature-informed indicators for near-time events.

\medskip
\emph{How to improve the computational efficiency and accuracy of a model?}
In order to gain online computational efficiency at the cost of offline simulations, a physical kernel approach can be imposed in training ML models to enable accurate but fast coarse scale simulation while still intelligently modeling the representation losses and residuals using deep neural networks (i.e., incorporating numerous functional forms of known closure kernels in the training as opposed to static phenomenological models\cite{maulik2018data,maulik2019sub,pawar2020priori,layton2020diagnostics}).
The training data can be also compressed by using both linear (e.g., matrix factorization\cite{gittens2016matrix}) and nonlinear techniques (e.g., variational autoencoders\cite{tait2020variational}, manifold learning\cite{maggioni2016multiscale,lu2019nonparametric,zhu2018image} and stochastic neighbor embedding algorithms\cite{hinton2002stochastic,maaten2008visualizing,linderman2019clustering}), and the causality assessment tools\cite{guo2008partial} can be employed to explore the correlation between input kernels and output variables in order to remove irrelevant input features from the training. These efforts can immensely reduce the size of the training data. The overall imposition of mass, momentum and energy conservation principles will be used as an inbuilt sanity check mechanism on the computations made by the neural network. Moreover, instability and inaccuracy resulting from the dimensionality reduction in the proposed HAM models can be corrected by a long-short term memory (LSTM) architecture as illustrated in our recent studies \cite{pawar2019deep,rahman2019nonintrusive}. To address transient characteristics, modal deformation and generalization of the ROM across a wide range of scenarios, it is possible to enhance HAM by utilizing principal interval decomposition \cite{san2015principal,ahmed2018stabilized,ahmed2019memory} and Grassmann manifold interpolation \cite{ahmed2020long,pawar2020data,pawar2020evolve}.
One can also utilize a \emph{method of manufactured learning} approach, which is primarily focused on devising statistical sampling and upscaling strategies to efficiently learn neural network architectures without requiring expensive simulations as training data. This approach has roots in the \emph{method of manufactured solutions} \cite{ROY2005131} (i.e., generating an auxiliary source function for a given chosen solution to validate the solver) that is often used to validate scientific codes. From this perspective, HAM will be an effective tool for the unbiased analysis of multi-x systems, especially with sparse datasets. The sparsity challenges can be handled by incorporating generative adversarial networks (GANs)\cite{goodfellow2014generative}. The method of manufactured learning can leverage both mathematical functions (e.g., Chebyshev, Dickens, wavelet families) or physical sets (e.g., proper orthogonal decomposition modes \cite{berkooz1993proper,liang2002proper,reiss2018shifted}, Lagrangian coherent structures \cite{shadden2005definition,haller2015lagrangian,hsieh2018small}, exact coherent structures \cite{waleffe2001exact,eckhardt2018small,beaume2014exact,suri2017forecasting}, terrain induced vortices \cite{thorpe2018application,restelli2009conservative,doyle2016numerical}, traveling waves \cite{james2013quadratic,hakkaev2017periodic,barker2020parameterization}, periodic or relative periodic orbits \cite{cieliebak2004symplectic}) to train $\mathcal{N}$ without requiring an expensive ``truth" labeled data. It is also pivotal when we construct self-evolving surrogate models on locally embedded structures. In principle, this approach can be exploited as a \emph{new training mechanism} for large data sets (e.g., geophysical turbulence) to address the generalizability and interpretability concerns of any ML approaches. 

\medskip
\emph{How to improve the trustworthiness of models? } In addition to enriching the input feature space as explained above, we can improve the trustworthiness of data-driven parts by imposing constraints and physical laws on the methods through the regularization of the cost function \cite{raissi2019physics} as well as by developing a scalable and optimal neural architecture search \cite{maulik2020recurrent} approach for model structure selection. To develop a more quantitative insight into the effects of these approaches, we can exploit the use of piece-wise affine representation of networks as shown in our recent work \cite{robinson2019ddn}. Moreover, a relatively straightforward approach for constructing error correction models based on formulating a data assimilation procedure can be proceeded in order to achieve long-term accurate prediction beyond training zone. To tackle stability issues for the online deployment of data-driven closures, it is possible to exploit the \emph{sequential data assimilation} techniques \cite{ahmed2020pyda} to correct the HAM predictions in Eq.~\ref{eq:main}. In particular, we can use a set of neural network architectures to learn the correlation between prognostic variable $\boldsymbol u$ and the data-driven parameterizations of $\mathcal{N}$, and formulate a \emph{nudging data assimilation} approach to improve long-range predictions in chaotic systems. The core idea behind nudging is to penalize the dynamical model evolution with the discrepancy between the model's predictions and observations \cite{anthes1974data,lei2015nudging,stauffer1993optimal}. In other words, the forward model given in Eq.~\ref{eq:main} is supplied with a nudging (or correction) term rewritten in the following discrete form, 
\textcolor{rev}{
\begin{equation}\label{eq:nudge1}
    \mathbf{u}^{n+1} = \mathbf{\mathfrak{M}}(\mathbf{u}^n) + \mathbf{G}(\mathbf{z}^{n+1}-h(\mathbf{u}_b^{n+1})), 
\end{equation}
where $\mathbf{\mathfrak{M}}$ is the forward model, $\mathbf{u}_b^{n+1}$ is the prior model prediction computed using imperfect background model, defined as $\mathbf{u}_b^{n+1} = \mathbf{\mathfrak{M}}(\mathbf{u}^n)$, $\mathbf{G}$ is called the nudging (gain) matrix, $\mathbf{z}$ is the set of measurements, and $n$ refers to the time index where we have these observations,} while $h(\cdot)$ is a mapping from model space to observation space. For example, $h(\cdot)$ can be a reconstruction map, from ROM space to FOM space as shown in our recent studies \cite{ahmed2020AIAA,ahmed2021nudged}. In other words, $h(\mathbf{u})$ represents the ``model forecast of the measured quantity'', while $\mathbf{z}$ is the ``actual'' observations. Despite the simplicity of Eq.~\ref{eq:nudge1}, the determination of the gain matrix $\mathbf{G}$ is not as simple \cite{zou1992optimal,vidard2003determination,auroux2005back,lakshmivarahan2013nudging}. Here, instead, we can define a functional nudging map $\mathbf{\mathfrak{C}}(\mathbf{u}, \mathbf{z})$ as,
\begin{equation}\label{eq:nudge2}
    \mathbf{u}^{n+1} = \mathbf{\mathfrak{M}}(\mathbf{u}^n) + \mathbf{\mathfrak{C}}(\mathbf{u}_b^{n+1}, \mathbf{z}^{n+1}), 
\end{equation}
where the map $\mathbf{\mathfrak{C}}(\mathbf{u}, \mathbf{z})$ will be inferred (e.g., LSTM model). 
We have tested this \emph{predictor-corrector nudging} methodology in our recent studies \cite{pawar2020data,pawar2020longPoF}. In a demonstration for solving the multiscale Lorenz 96 system\cite{lorenz1996predictability}, the chaos contaminates the solution in a short time when $\mathcal{N}$ is solely modeled by a neural network to account for the fast fluctuating dynamics. However, we showed a significant improvement in the long-term prediction of the underlying chaotic dynamics with our forecast error correction framework empowered by a deterministic ensemble Kalman filer (DEnKF) \cite{sakov2008deterministic}. \textcolor{rev}{For more detailed discussions on algorithms of such nonlinear filtering frameworks, we refer readers to our recent comparative study of the LSTM nudging and other sequential data assimilation methods for the Kuramoto-Sivashinsky equation \cite{pawar2021comparative}.} Moreover, we demonstrated that the HAM approach can handle the \emph{non-Gaussian statistics} of subgrid scale processes, and effectively improve the accuracy of outer data assimilation workflow loops in a modular non-intrusive way. While these preliminary results indicate the basic viability of the approach, they do not cover many key aspects that need to be addressed to scale this predictor-corrector nudging idea into more realistic situations. It is also interesting to note that the similarities and equivalence between ML and the formulation of statistical data assimilation as used widely in physical and biological sciences have been postulated \cite{geer2020technical,abarbanel2018machine}.

\medskip
\emph{How to make generalizable models?} 
%We will build on our recent works in reinforcement learning (RL) architectures \cite{meyer2020taming,theie2020deep,meyer2020colreg}, in which we demonstrate that RL agents are able generalize to more realistic scenarios despite being trained in a very synthetic environment. 
The problem of building a model from the data can be posed as a multi-objective optimization problem where several aspects of the knowledge-injection processes must be involved in data preparation, feature engineering, training and testing steps.
One can investigate how the reinforcement learning (RL) agent trained with an environment with the coarse mesh behaves in the environment with fine resolution. These efforts can be built on our recent works in RL architectures \cite{meyer2020taming,theie2020deep,meyer2020colreg} to devise new approaches for transfer learning from a low-fidelity model to a high-fidelity model. 
It would be a nice attempt to explore constructing new reward functions that are more suited to multi-x environments to take into account the efficient sampling of big data. A \emph{design of experiments} methodology can be used to keep the number of numerical experiments (required to develop HAM) as small as possible without compromising the resulting quality. The agent will then be trained with the new objective of adaptive sampling. Moreover, to put the learning of the agent in the form of ordinary differential equations one can exploit the use of a \emph{symbolic regression} method for hidden physics discovery \cite{vaddireddy2020fes}. A symbolic regression approach built on \emph{gene expression programming} \cite{schmidt2009distilling} and \emph{graph networks} \cite{cranmer2020discovering} has an edge over other ML methods in the sense that it gives a functional form of the mapping from the input to the output space making stability analysis for the system possible. With these augmented RL agents one can synthesize data and take advantage of the expressivity of symbolic and multivariate regression tools to discover new parameterizations. As an alternative approach, we can take advantage of \emph{equation-free multiscale methods} \cite{kevrekidis2003equation,sirisup2005equation,kevrekidis2009equation} that allow us to compute a system on a level for which no complete governing equation exists. This coarse graining approach (a.k.a. \emph{projective integration}) requires identification of the latent space that represents the system without ever deriving it in a closed form. The success of this approach often depends upon the existence of a separation of time scales on the system dynamics. 
To determine when the errors introduced by the coarse projective steps have healed, and therefore another projective step can be taken, one can monitor, for example, energy spectra or structure functions in fluid dynamics applications. This will provide us a sanity check mechanism for the spatial, temporal and internal consistency of the ML outputs. 

\medskip
\emph{How to develop self-evolving models which learn and adapt as-we-go?}
This can be accomplished by blending new measurements through dynamic data assimilation to correct and update HAM models. While one can continuously and recursively update the reduced dimensional basis with the most recent observations, the models can be made to adapt to recent changes while still exploiting knowledge learned a long time in the past. To match the output of the hybrid model with measurement data as-we-go, one of the key aspects in accelerating overall HAM workflow is to create a robust data assimilation module that can also be instrumental in sensitivity, robustness and error propagation analysis. Moreover, the resulting hybrid model can be applied (i) as an inner-engine with the data being observations streamed in through the outer-loop; and (ii) as a surrogate for the linear solution process within physics-based nonlinear solvers, where now the data stream is provided by the nonlinear solution process itself. The new perturbations to the system will be captured using the sparse data collected by the sensors to further develop our evolve-then-correct reduced order modeling strategy \cite{pawar2020evolve}. Invariably, numerical approximations of transient and nonlinear phenomena can lead to sequences of nonlinear algebraic systems. 
At the start of a simulation, an offline HAM can act as a rough preconditioner for the linear solution process. As the simulation progresses, the nonlinear iterates might act as the online data-stream that the HAM assimilates, thereby creating a symbiosis. As the HAM continues to learn and adapt, it can ultimately become as an efficient preconditioner. 
Furthermore, low-rank approximations can be injected into a numerical integration itself. Therefore, the target algebraic system remains to be the full scale problem, while the nonlinear solution process itself exploits a ROM representation.

\medskip
\emph{How to better quantify uncertainties?}
%\hl{Mention the importance of uncertainty quantificiation (some inspiration from the RAPDI proposal)}
The decision-making is a complex process for many disciplines. In recent decades, computational models have shown the potential of supporting the decision, providing detailed information that may not be directly acquirable in experiments. However, the sensitivity of the results on the different sources of uncertainties is a critical piece of information. Uncertainty quantification (UQ) therefore certifies the reliability of computational
models working with noisy data. However, a UQ analysis often requires many simulations to estimate the unknown coefficients (e.g., polynomial chaos expansion coefficients \cite{crestaux2009polynomial} with a set of quadrature points). We can use reduced order models and data-driven tools to improve the efficiency of the underlying decision-making process by performing a complete UQ analysis in a reduced amount of time. We envision that the HAM approach can significantly accelerate the UQ analysis to feasible CPU times (e.g., on the order of minutes) on local computers. This becomes even more important in certain fields (i.e., health care and biomedical applications), where the high performance computing (HPC) facilities might not be easily accessible for parallel computations. Moreover, sensitive data-protection issues can make remote HPC simulations very challenging.  

\medskip
The above mentioned challenges in each question are somehow interrelated, so progress in each front must build on advances in the other ones. Generally speaking, the black-box parameterization often becomes an ill-posed problem and might diverge when considering long duration integrations. There are several remedies already available. Namely, one can explore including the mean Jacobian approximation \cite{chevallier2001evaluation}, weight smoothing \cite{aires1999weight} or other regularization techniques, and cost function modifications \cite{lee1997hybrid}, where the first order Sobolev norm has been used in the training instead of the standard Euclidean norm. When deeper neural network architectures are required, the complexity might become a practical issue, e.g., in the GCM computations, since the learning of the weights is very time-consuming and requires a large optimization problem that is prone to yield sub-optimal solutions. To mitigate this one can utilize the \emph{extreme learning machine} closure approach \cite{maulik2017neural} in a recursive manner for an easier training process (e.g., for re-training a part of $\mathcal{N}$ to adapt it to the new environment).  

\medskip
\textcolor{rev}{
To sum up, bringing physical realism in DTs of next-generation physical and engineering systems will need (i) new modeling approaches that are accurate and reliable, generalizable, computationally efficient, and trustworthy, and (ii) seamless integration of multi-scale, multi-physics, and multi-fidelity models (multi-x systems). To this end, we will highlight two HAM approaches in the following sections that address the above needs. While the first approach, physics-guided ML, provides a mechanism to inform the black-box ML models like a deep neural network in which the first principles feed the training process to enhance the predictive model's trustworthiness, the second approach, interface learning, enables the coupling of multi-x systems. 
}

\begin{figure}[ht]
\centering
\includegraphics[trim= 0 0 0 0, clip, width=0.95\textwidth]{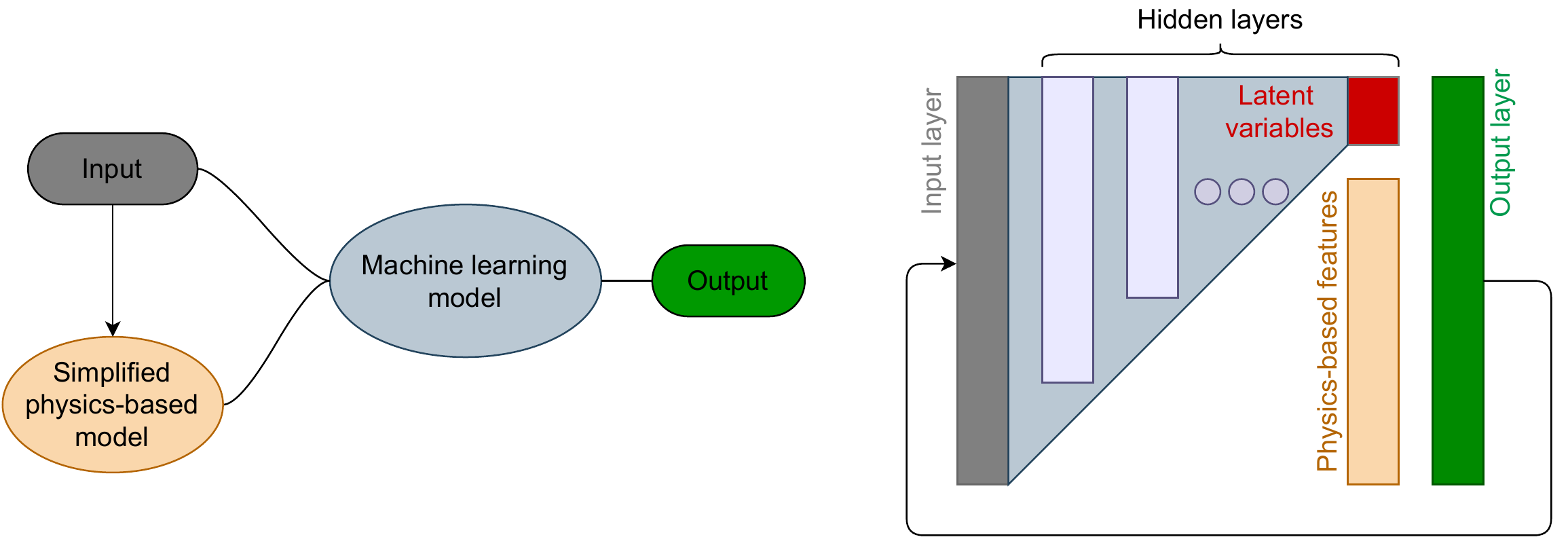}
\caption{\textcolor{rev}{A physics-guided ML framework, where the physics-based features are directly embedded into hidden layers of the neural network. As an illustrative example to this framework, we can consider constructing a learning engine to obtain a surrogate model in replacing CFD calculations for the aerodynamic load prediction tasks in demand. Input layer could be an airfoil shape, and output layer could be the lift and drag coefficients. We can construct our ML model to incorporate physics into ML models through embedding simplified theories (e.g., panel method as a fast physics-based predictive model) directly into neural network models.}}
\label{fig:pgml}
\end{figure}

%\section{Eclecticism}\label{sec:ec}
\textcolor{rev}{
\section{Neurophysical modeling and physics-guided ML}\label{sec:pgml}
There is a growing interest in applying ML, particularly deep learning for scientific applications due to the abundance of data generated either from high-fidelity numerical simulations or sparse measurements. Despite the success of data-driven models for many problems in scientific computing, they lack the interpretability of the physics-based models. This motivates the need to address the generalizability and data efficiency of the data-driven approaches for physical systems. Therefore, one of the active research thrusts is to leverage methodologies for the combination of physics-based and neural network models, a rapidly emerging field that came to be known as physics-guided AI (PGAI) or physics-guided ML (PGML). To this end, we have recently introduced a PGML framework where information from simplified physics-based models is incorporated within neural network architectures to improve the generalizability of data-driven models \cite{pawar2021physics}. The central idea in the PGML framework is to embed the knowledge from simplified theories directly into an intermediate layer of the neural network as shown in Figure~\ref{fig:pgml}. The knowledge from the simplified theories aids in ensuring that we learn only the knowledge required to compensate for the deficiencies of these theories instead of learning everything from scratch. Also owing to the fact that the simplified theories are still based on the laws of nature they are more generalizable compared to any data-driven approach and hence they should be exploited to the extent possible. For example, the prediction of flow around an airfoil is a high-dimensional, and nonlinear problem that can be solved using high-fidelity methods like computational fluid dynamics (CFD). However, a wide variety of problems such as aerodynamic performance prediction \cite{zha2007high,legresley2000airfoil} for optimal control, design optimization, uncertainty quantification might require the prediction of the quantity of interest in real-time, and these CFD methods become computationally infeasible for these real-time prediction purposes. To overcome these challenges, utilizing ML approaches to build a data-driven surrogate model is gaining widespread popularity \cite{zhang2018application,bhatnagar2019prediction}. Yet one of the well-known issues with neural networks is that these universal approximators exhibit poor \emph{generalizability}, i.e., they produce inaccurate prediction when the test data is from a distribution far from the training data. This adversely affects the trustworthiness of neural networks for scientific applications. Therefore, it is important to incorporate the domain knowledge into learning, and one can exploit the relevant physics-based generalizable features from simplified models (e.g., panel methods) through the PGML framework to enhance the generalizability of data-driven surrogate model. }

\textcolor{rev}{
In a typical neural network architecture, the input vector $\mathbf{x} \in \mathbb{R}^m$ is fed to the supervised ML model, and the mapping from the input vector to the output vector $\mathbf{y} \in \mathbb{R}^n$ is learned through training. The neural network is trained to learn the function $F_\theta$, parameterized by $\theta$, that includes the weights and biases of each neuron. The parameters of the neural network are optimized using the backpropagation algorithm to minimize the cost function. Usually, for the regression problems, the cost function is the mean squared error between true and predicted output, i.e., $C(\mathbf{x},\theta)=||\mathbf{y} -F_{\theta}(\mathbf{x})||_2$. In the PGML framework \cite{pawar2021physics}, the neural network can be augmented with the output of the simplified physics-based model. The features extracted from simplified physics-based models are embedded into hidden layers along with latent variable. This is in contrast to admitting physics-based features at the input layer in conventional neural network architectures, which might lead to underestimation of the effect of such physical information, especially for high dimensional systems. For example, imagine stacking a few parameters to an input vector of a dimension of $O(10^3-10^6)$ (e.g., coordinates describing the shape of an airfoil or blade). It is highly probable that the learning algorithm overlooks the effect of such prior knowledge (scalars like the angle of attack and Reynolds number of the flow) in the minimization algorithm. In the PGML, a proper latent space is first identified and the information from the physics-based model is injected into the neural network. Since the neural network is no more required to learn these features by itself its complexity and associated model uncertainty can be significantly reduced.   
}

\textcolor{rev}{
As demonstrated in predicting aerodynamic loads \cite{pawar2021physics}, the PGML framework is successful in significantly reducing the model uncertainty (captured using the prediction from multiple models initialized with different parameters) of the neural network. It has been shown that the PGML framework also enhances the generalizability of the neural network based surrogate model since it predicts the quantity of interests more accurately for unseen conditions. We highlight that the PGML framework can be applied in different branches of science and engineering, such as wind farm layout optimization, boundary layer flows, where simplistic models are very common. The simplified models are cheap to run and therefore the PGML framework can be attractive for applications such as model predictive control, data assimilation, and uncertainty quantification. The hybrid model built using the PGML framework can act as an enabler for predictive simulations paving the way for the DT of physical systems. In fact PGML can be combined with PINN \cite{raissi2019physics,zhu2019physics,pan2020physics} to exploit the latter's full potential. In PINN, the cost function is regularized using the residual of physics-based equations. The regularization can make the task of minimizing the cost function difficult specially when the complexity of neural network is very high. By utilizing PGML one can reduce the complexity of the model thereby making the optimization process more efficient.   
}

\section{Eclecticism and interface learning}\label{sec:il}
%\hl{Multiphysics, Multifidelity Integration of DT at component levels to build DT at an asset level}
In this section, we focus on the eclectic interface modeling framework for multi-x systems. The HAM approaches can offer the key enablers to develop multifidelity formulations that consist of multiple abstractions with different characteristics. Unlike traditional iterative schemes, first of all one can develop and explore the utility of a spectrum of HAM methodologies for non-iterative interface learning. To expedite computations, different levels of models and descriptions should be devoted to different zones and components of the problem in order to allocate computational resources more effectively and economically. This might be the case for many other coupled multiphysics systems relevant to fluid dynamics, such as heterogeneous multiscale problems\cite{kevrekidis2009equation,weinan2003heterogeneous, quarteroni2016geometric, quarteroni2003analysis,hughes2000large,hughes2001multiscale}, fluid-structure interaction \cite{van2011partitioned, NORDANGER2016324}, aerospatial \cite{shankaran2001multi,cook2019optimization,xu2020reduced}, subsurface \cite{bjorkevoll2015use,macpherson2015technology}, and wind farm applications \cite{siddiqui2019numerical}. Since various zones in these systems are connected through interfaces, data sharing and consistent interface treatment among respective models are inevitable (e.g., Dirichlet-Neumann, Dirichlet-Robin, and Robin-Robin). Likewise, multirate and locally adaptive stepping methods can yield a mismatch at the space-time interface, and simple interpolation or extrapolation might lead to solution divergence or instabilities \cite{gander2013techniques}. Meanwhile, even if we are interested in simulating just one portion of the domain corresponding to specific dynamics, we still need to specify accurate interface conditions. Running a high fidelity solver only to provide the flow state at the interface seems to be unreasonable. Recently, we formulated an eclectic interface modeling approach as a key enabler for emerging DT technologies in many sectors \cite{rasheed2020digital}. However, just like any technology, it comes with its own needs and challenges \cite{tao2018digital,hartmann2018model,ganguli2020digital,chakraborty2020machine,kapteyn2020data,kapteyn2020physics}. 
Therefore, an accurate characterization and modeling of the interface is crucial to derive the consistent boundary conditions and upscaling laws in a large variety of scientific applications.
\textcolor{rev}{As discussed in our recent work \cite{ahmed2020interface}, complex natural or engineered systems often comprise multiple characteristic scales, multiple spatiotemporal domains, and even multiple physical closure laws. To address such challenges, an \emph{interface learning} (IL) paradigm, which consists of a data-driven closure approach based on memory embedding, has been introduced to provide physically consistent boundary conditions at the interface. In a nutshell, here we can define IL as the utilization of statistical tools to provide some \emph{educated} or \emph{learned} interface conditions to couple different solvers or models to maximize computational efficiency. Although interface modeling approaches are also possible with classical physics-based approaches (e.g., see \cite{pawar2020lbm}), we highlight that HPC environments can benefit from this methodology to reduce communication costs among processing units in emerging machine-learning-ready heterogeneous platforms toward exascale era.} Moreover, it is possible to include physical insights into an interface learning framework, like \emph{upwind learning} concept \cite{ahmed2020interface} that can be considered as a physics-informed data-driven domain decomposition approach. The chief idea has been drawn to enforce physics into the learning process. This process, therefore, accounts for the effects of downstream domain with an ML model that provides the interface condition in order to be able to solve for the upstream domain. Another key aspect of such a zonal multifidelity approach is its ability to handle intrinsic heterogeneous physical properties, varying geometries, and underlying governing dynamics. This heterogeneity can be pronounced in certain applications with spatially varying states. Many interfacial problems can be put into the following categories, explained with a family of examples as follows:

\medskip
\emph{Reduced order model - full order model (ROM-FOM) coupling:} 
As we highlight in our recent work \cite{rasheed2020digital}, there is a demand for lighter models that can run in real-time. 
In the context of surface flows (e.g., urban modeling and weather prediction), FOMs have been in use for a long time; however, they are incapable of modeling phenomena associated with scales smaller than what the coarse mesh can handle (like buildings and small terrain variations). These fine scale flow structures can be modeled using a much refined mesh but then the simulations become computationally intractable. To tackle this problem one can systematically couple FOM and ROM models, and develop consistent interface conditions (both in space and time) using, for example, the nudging type approaches introduced in Section~\ref{sec:ham}. In setting over complex terrains, the ROM-FOM coupling could be crucial, where we can utilize a high-fidelity multi-zone large eddy simulation (LES) solver to simulate flows in complex terrains, which are dominated by patterns like channeling, rotors, hydraulic jumps, and mountain waves \cite{siddiqui2019numerical}. Once we build ROMs, we can develop new coupled ROM-FOM models to perform a real-time, accurate approximation of wind turbine loadings due to wind flow interactions, which is essential in maximizing the power production of wind farms. Several methods to bridge low-fidelity and high-fidelity descriptions have been recently introduced to form the building blocks of an integrated HAM approach among mixed fidelity descriptions \cite{ahmed2021multifidelity}. 

\medskip
\emph{Multiphysics and multiscale coupling:} Various dichotomies with a multiphysics coupling of interacting subsystems can be identified and exploited in a multitude of scientific applications. A common example in this category is the use of a particle based approach (e.g., molecular dynamics) in part of the domain, while using a continuum based approach (e.g., Navier-Stokes) in the rest of the domain \cite{o1995molecular,mohamed2010review,cosden2013hybrid}. In a recent study\cite{pawar2020lbm}, the feasibility of such an interface learning approach has been demonstrated by solving the FitzHugh-Nagumo model in a bifidelity setting partitioned between a finite difference method (FDM) and the lattice Boltzmann method (LBM), where the authors consider the underlying dynamics to be composed of a collection of pseudo-particles that are represented by a particle distribution function. 

\medskip
\emph{Geometric multiscale or mixed-dimension modeling:} It is desirable to develop mathematical models for physical problems in which it is necessary to simultaneously consider equations in different dimensions \cite{d2008coupling,boon2018robust,nordbotten2019unified,quarteroni2003analysis,passerini20093d,quarteroni2016geometric}. For example, a network flow \cite{ahmed2020interface} or a Darcy subsurface flow system \cite{sheth2017localized} can be mathematically formulated by means of heterogeneous problems featuring different degrees of detail and different geometric dimensions that interact together through appropriate interface coupling conditions.

\medskip
\emph{Model fusion:} From a fluid dynamics perspective, there is a plethora of models to describe subgrid scale closures or nonlinear residual effects. Turbulence modeling generally requires an apriori selection of the most suited model to handle a particular kind of flow. However, it is seldom that one model is sufficient for different zones in the computational domain. To alleviate this problem, hybrid and blending models have been extensively utilized to lift technical barriers in industrial applications, especially in settings where the RANS approach is not sufficient and LES is expensive  \cite{sagaut2013multiscale,fadai2010seamless,shur2008hybrid}. By utilizing ML classifiers \cite{maulik2019sub,maulik2020spatiotemporally} and regressors \cite{maulik2019subgrid,maulik2017neural,beck2019deep}, we can develop hybrid RANS/LES turbulence models to provide the exchange of information accurately at the interface, and explore new hybridization criteria.

\medskip
\emph{Nested solvers:} To decrease the computational cost required for an accurate representation of the numerous interconnected physical systems, e.g., oceanic and atmospheric flows, it is pivotal to develop several classes of nested models to form the basis of highly successful applications and research at numerous weather and climate centers \cite{waldron1996sensitivity,von2000spectral,radu2008spectral,miguez2004spectral,rockel2008dynamical,schubert2017optimal}. Numerous HAM tools can be used to force the large-scale atmospheric states from global climate models onto a regional forecasting model, e.g., see a canonical example in demonstrating an LSTM-nudging approach \cite{pawar2020longPoF}.

\medskip
\emph{Space-time domain decomposition:} An analogous situation usually occurs in parallel computing environments with domain decomposition and distribution over separate processors with a message passing interface to communicate information between processors. The heterogeneity of different processing units creates an asynchronous computational environment, and the slowest processors will control the computational speed unless load-balancing is performed \cite{donzis2014asynchronous,mittal2017proxy}. The feasibility of such space-time techniques toward noniterative domain decomposition has been demonstrated by introducing the \emph{Learn from Past} (LP) concept \citep{ahmed2020interface}. Since the interface neighboring points might be evolved in time before the interface condition is updated in an explicit scheme, a variant of the LP model based on a combination between old and updated values, namely the \emph{Learn from Past and Present} (LPP) model \citep{ahmed2020interface}, can be utilized as well. It is possible to extend this mapping to take into account the time history dependence in a non-Markovian manner through the adoption of the recurrent neural networks (RNNs).  

%\section{Multifidelity Computing}\label{sec:multifidelity}
\medskip
\textcolor{rev}{
\emph{Interfacing of DTs:} DTs can be developed for different components constituting an asset. For example in the context of wind farm there can be component level DTs of turbine blades, drive trains, atmospheric condition, energy trading. Interface learning can once again facilitate bidirectional exchange of information between these component level DTs to build a holistic asset level (e.g., wind farm) DT.}  

\textcolor{rev}{
\section{Reduced order modeling, data assimilation and control}\label{sec:rom}
A chief motivation behind the reduced order modeling is to be able to use them in multi-query applications such as control and optimization. Given the abundance of data and data-driven tools, the development of ROMs is gaining increasing popularity nowadays. As highlighted in our introduction, there are a great number of review articles available concerning various aspects of model order reduction and its applications \cite{kashinath2021physics,bauer2021digital,geer2021learning,brunton2020machine,taira2020modal,frank2020machine,zhu2020river,tahmasebi2020machine,willard2020integrating,mendoncca2019model,yu2019non,brenner2019perspective,yondo2019review,taira2017modal,rowley2017model,benner2015survey,brunton2015closed,asher2015review,lassila2014model,machairas2014algorithms,mezic2013analysis,mignolet2013review,bazaz2012review,razavi2012review,theodoropoulos2011optimisation,chinesta2011short,wagner2010model,massarotti2010reduced,kleijnen2009kriging,pinnau2008model,lucia2004reduced,ong2003evolutionary,freund2003model,bai2002krylov,chatterjee2000introduction,berkooz1993proper,bonvin1982unified,elrazaz1981review}, and more relevant to our discussion, the enabling role of model order reduction approaches in developing next generation DT systems has been also discussed \cite{hartmann2018model,rasheed2020digital}. Of particular interest, model predictive control (MPC) \cite{garcia1989model,mayne2014model,hewing2020learning} originated in the late seventies and has since then evolved considerably. It is not a specific control strategy but rather a range of control methods which make explicit use of linear and even nonlinear models to obtain the control signal by minimizing an objective function over a future horizon. One of the keys to success of these models is a continuous update of the horizon and computing an optimal control input while anticipating the future state. It is clear that the computation of an optimal control input thus depends heavily on the accuracy and computational efficiency of the models. Unfortunately most of the realistic models are highly non-linear and computationally stiff to solve making the optimization process a bottleneck in many of the MPC processes. One approach that has been extensively used is to linearize the model around a set point and then use the traditional linear control theories \cite{muske1993model}, which are founded on the basis of the superposition principle. As one moves from linear to nonlinear systems these principles no longer hold and suffers from two major issues: firstly, it cannot predict the nonlocal behavior far from the operating point (e.g., long horizons) and secondly it often ignores nonlinear phenomena which can take place only in the presence of nonlinearities \cite{grune2017nonlinear}. To this end, reduced order modeling offers an attractive alternative when designing a digital cyberinfrastructure. Overall, ROM can play four major roles in the context of MPCs:
\begin{itemize}
    \item[i.] \textit{Data compression:} A ROM approach can be used to compress the data from a system. This can be useful when the communication bandwidth is limited and storage space is expensive. Furthermore, instead of saving the ROM rather than the data can make retrieval or archival of data really fast, something desirable when a continuous update of the ROM is required. 
    \item[ii.] \textit{Deducing quantity of interest:} Even with the ability to collect big data through the deployment of multiple sensors in a system, it might be difficult to compute the quantity of interest. Complex models might be required for such computation. ROM in this regard can help in speeding up such modelling while still exploiting the information available in big data \cite{kapteyn2020probabilistic}.
    \item[iii.] \textit{Accelerating the optimization process:} The optimization process requires running the model repeatedly to find out optimal control parameters over a prediction horizon. ROMs which can give accuracy comparable to the high fidelity models but still with several order of magnitude reduction of computational cost \cite{queipo2005surrogate} can make such practical even when the dynamics is nonlinear. 
    \item[iv.] \textit{Accelerating stochastic modeling:} Probabilistic modeling often includes Monte Carlo simulations which are costly using the FOM. However, in a recent study~\cite{farhat2018stochastic} the authors show how ROM together with hyperreduction can be used to speed up stochastic modeling to take into account model-uncertainty in the underlying physical model.   
\end{itemize}
We envision that ROMs will be a key enabler to lay the groundwork for developing a big data cybernetics infrastructure (e.g., please see Section~\ref{sec:big}), an approach to controlling an asset or process using real-time big data. These tools and concepts offer many new perspectives to our rapidly digitized society and its seamless interactions with many different fields. With the recent wave of digitalization, the latest trend in every industry is to build systems and approaches that will help it not only during the conceptualization, prototyping, testing and design optimization phase but also during the operation phase with the ultimate aim to use them throughout the whole product life cycle and perhaps much further beyond that. While in the first phase, the importance of numerical simulation tools and lab-scale experiments is not deniable, in the operational phase, the potential of real-time availability of data is opening up new avenues for monitoring and improving operations throughout the life cycle of a product, and ROMs will be crucial to improve such DT technologies. Such a holistic approach can find application in many fields such as wind energy for optimal control and predictive maintenance, in autonomous vehicles for optimal and safe maneuvering under adverse environmental conditions, in drug delivery, and beyond.} 

\section{Big data cybernetics}\label{sec:big}

\begin{figure}[ht]
	\centering
	\includegraphics[width=0.8\textwidth]{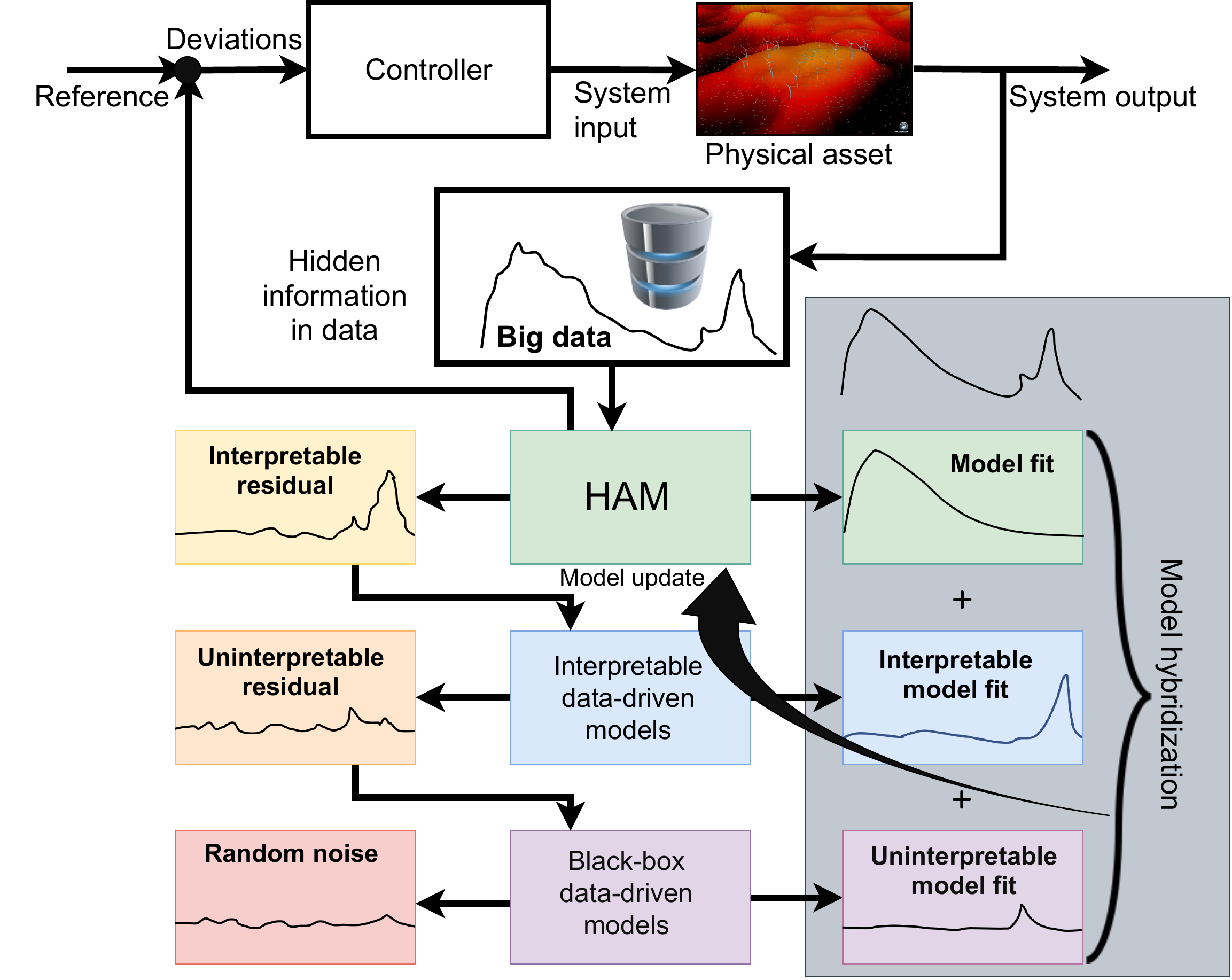}
	\caption{An overview of big data cybernetics: The model starts as a physics-based model that is based on first principles but as time progresses the model is continuously updated using knowledge generated from data. The figure is adapted from Rasheed et al. \cite{rasheed2020digital}.} 
	\label{fig:big}
\end{figure}

By combining methods from control theory and HAM, big data cybernetics is an approach to controlling an asset or process using real-time big data. As an explainable and interpretable approach envisioned by Martens\cite{martens2015quantitative}, it is aimed to discover systematic/coherent structures and convert them into more quantitative information. Although the big data cybernetics has been emerged as an approach without involving black-box tools such as neural networks or support vector machines, we envision the utilization of the HAM methodologies to process and translate big data into smart data. \textcolor{rev}{Using the specific needs of the industry, application, or process, big data often gets turned into smart data when it is optimized \cite{kalinin2015big,iafrate2015big,garcia2019enabling}. Here, we envision that the feature engineering and optimal sensor placement methodologies might serve as viable tools for this transformation.} In the context of upcoming technologies \textcolor{rev}{like DTs,} the role of cybernetics is to steer the system toward an optimal set-point as shown in Figure~\ref{fig:big}. The difference, called the error signal, is applied as feedback to the controller, which generates a system input to bring the output set-point closer to the reference. With the availability of more and more sensors and communication technologies, an increasingly larger volume of data (in fact big data) is being made available in real-time. At the first step, we interpret the big data using well understood mechanistic models based on known physics. We term the rest as interpretable residual that can be modeled using a data-driven approach. After this second step, an uninterpretable residual remains to be modeled using more complex and black-box models like deep neural networks. The remaining residual is generally noise, which can be discarded. The three steps result in a better understanding of the data and hence an improved mechanistic model resulting from the hybridization of physics-based and data-driven models, and will be continuously looped with the availability of new sources of data. This concept offers many new perspectives to our rapidly digitized society and its seamless interactions with many different fields.

We emphasize that new situations (e.g., situations that are not included in a supervised learning machine) are pervasive in scientific applications, especially in control problems. In many cases, there is a grand challenge in making accurate predictions using data-driven methods based strictly on statistical inference. Moreover, with partial or no understanding of the new situation, it might be almost impossible to develop a pure physics-based model due to lack of parameterizations from \textcolor{rev}{first principles.} Therefore, hybrid methods could be instrumental in discovering new control laws. For example, recently, we show how a reinforcement learning (RL) agent can learn complicated control laws through trial and error to achieve complicated tasks of path following and collision avoidance simultaneously \cite{meyer2020taming,theie2020deep,meyer2020colreg}. However, such generalizable learning happens in a black-box manner so their applicability in critical applications is foreseen to be limited unless the learned control laws can be expressed in comprehensible mathematical form. In this context, the developments of sparse regression and symbolic regression methods are highly encouraging \cite{schmidt2009distilling,brunton2016discovering,cranmer2020discovering}. In particular, our recent work exploited \emph{symbolic regression} (SR) as a powerful technique to discover highly non-linear equations directly from sparse sensor data\cite{vaddireddy2020feature}. A remaining question is how to combine RL and SR for making generalizable models without compromising interpretability. We envision that one can make use of open source RL packages to scale the training in a distributed manner by running multiple environments concurrently to accelerate the training. 
An exploration for constructing new reward functions that are more suited to certain multi-x environments to take into account efficient sampling of big data. A \emph{design of experiments} methodology can be used to keep the number of numerical experiments (e.g., required to develop HAM) as small as possible without compromising the resulting quality. The agent can then be trained with the new objective of adaptive sampling.

Computational models are nowadays implemented in many decision making processes like model predictive control \cite{camacho2013model,kaiser2018sparse}. For instance, flow control deals with altering the flow field to achieve the desired objective and is of immense importance in practical applications such as aerodynamic drag reduction, mixing enhancement, and noise reduction \cite{gad1996modern}. The problem of closed-loop control in fluid mechanics is challenging due to strong nonlinearity, high-dimensionality, and recent advances in system identification, model-order reduction, ML, and compressive sensing have made significant progress in adaptive real-time control \cite{brunton2015closed, noack2011reduced}. Unsupervised ML algorithms, like RL, have illustrated the potential of human-level control in playing games \cite{mnih2015human}, robot motion control \cite{zhang2015towards}, and for continuous control tasks \cite{duan2016benchmarking}. RL controllers are based on experience gained from self-play or exploration, using algorithms that can learn to execute tasks by reinforcing good actions based on a performance metric. The main framework of the RL consists of an agent (for example, a neural network in deep RL) that interacts with an environment to learn a policy that will maximize the cumulative reward over a long time horizon \cite{sutton2018reinforcement}. In recent years, the RL has been explored for fluid dynamics problems including animal locomotion \cite{gazzola2014reinforcement,verma2018efficient,reddy2016learning}, control of chaotic dynamics \cite{bucci2019control,beintema2020controlling,vashishtha2020restoring}, drag reduction of bluff bodies \cite{rabault2019,ren2020applying,tokarev2020deep}, flow separation control \cite{shimomura2020closed}, and turbulence closure modeling \cite{novati2020automating}. Along with a computer simulation environment, RL has been effectively applied for active flow control around bluff bodies in an experimental setup \cite{fan2020reinforcement}. Also, Novati et al. \cite{novati2020automating} introduced the multi-agent RL strategy to learn the turbulence closure as a control policy enacted by cooperating agents, and all the agents are trained to minimize the distance from direct numerical simulation (DNS) energy spectrum. Fan et al. \cite{fan2020reinforcement} utilized the actual towing experiment with three cylinders as an environment of the RL framework. In their work, the rotation of two cylinders in the wake of the main cylinder was used as an action to minimize the overall drag of the system, and they demonstrated the importance of noise reduction strategies like the Kalman filter for applying deep RL to experimental environments. Bucci et al. \cite{bucci2019control} applied the deep RL to build a model-free controller to stabilize a chaotic system using limited sensor measurements as observable by an agent. In another work, Ma et al. \cite{ma2018fluid} utilized RL to control a coupled system involving both fluid and rigid bodies, by applying control forces only at the simulation domain boundaries. Genetic programming (GP) has also  attracted great attention for active flow control frameworks. For example, GP was applied to control the recirculation area of a backward facing step \cite{gautier2015closed} and mitigate the separation and early reattachment of turbulent boundary layer for a sharp-edge ramp \cite{debien2016closed}. Moreover, Ren et al. \cite{ren2019active} adopted GP to select explicit control laws for the suppression of vortex-induced vibration of a circular cylinder in a low-Reynolds-number flow using blowing/suction at fixed locations. All these ideas can be exploited in emphasizing the potential for using flexible data assimilation and automatic control environments to unlock the capabilities of a DT framework.

\section{Digital twin revolution}\label{sec:dt}

\begin{figure}[ht]
	\centering
	\includegraphics[width=\textwidth]{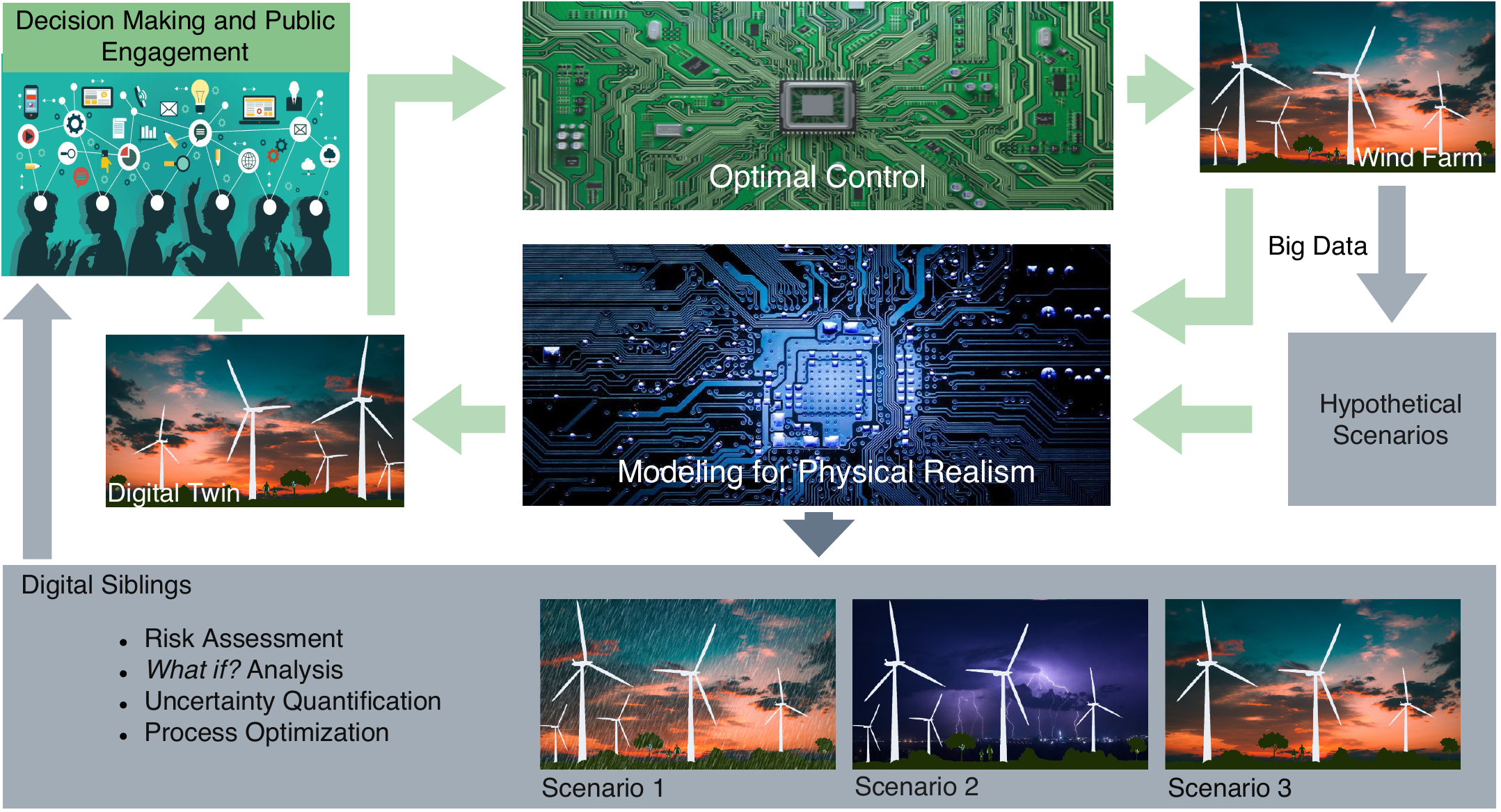}
	\caption{An overview of a DT concept where the HAM within the big data cybernetics framework becomes a key enabler for modeling physical realism and optimal control. For instance, a DT of the wind farm can allow us to evaluate different scenarios using hybrid models. The big data collected from IoT sensors can be continuously assimilated to correct hybrid models and improve the state estimation.} 
	\label{fig:DT}
\end{figure}

We can arguably discuss that the social media is the most recent disruptive technology for generating DTs of users. Why not generating DTs of assets or processes? Therefore, in a broad sense, we can define a DT as a virtual representation of an asset or its components for real-time prediction, monitoring, control, and optimization of the asset throughout its life cycle. This DT concept with a high level of realism has been started to be a paradigm shift for more informed decision making in every industry. \textcolor{rev}{It is quite natural that improving preconditions from simulations and models will lead to an improvement of DT technologies. Hence, the question arises why HAM techniques specifically could enable the revolution of DT technologies.}  In our recent survey paper\cite{rasheed2020digital}, we have presented a DT framework and detailed its values, challenges and enabling technologies.
\textcolor{rev}{
For example, some of the major challenges in realizing the potential of wind energy to meet the global electricity demand are the need for a deeper understanding of the physics of the atmospheric flow, science, and engineering of these large dynamic rotating machines and synergistic optimization and control of fleets of wind farms within the electricity grid \cite{veers2019grand}. The decades of research and development in fluid dynamics, systems engineering, numerical methods, manufacturing processes, and material discovery can now be complemented with unprecedented amounts of data generated from insitu measurements, lab experiments, and numerical simulations to tackle these challenges. As we discussed in Section 2, the combination of physics-based and data-driven models is increasing in every branch of science leading to numerous novel HAM approaches for many scientific applications. Synergistically, the HAM paradigm can be applied to a variety of tasks related to wind energy research, such as DTning for the optimization and real-time control of wind farms. A DT of the wind farm lets us examine \emph{what if} scenarios, evaluate the system's response, and select the corresponding mitigation strategies. As illustrated in Figure~\ref{fig:DT}, the DT of wind farms can be useful for different purposes such as optimal control of wind turbines to achieve the maximum performance, routine maintenance of the equipment, accurate forecast of the power production, wind farm optimization \cite{barter2020systems}, and improved decision-making. The success of the DTs depends upon the type of approach that we employ for modeling the system. The ability of the HAM to combine the generalizability of physics-based approaches and the automatic pattern-identification feature of data-driven approaches makes it an attractive choice to model virtual replica of physical systems in DTs. For example, there are several turbine wake models that capture the flow in the wake of a single turbine and these models predict sufficiently accurate aerodynamics of a single turbine \cite{bastankhah2014new,shapiro2018modelling}. However, these models are not enough to capture the flows in wind farm due to wake superposition, complex terrains, deep array effects, and neglected physics \cite{meneveau2019big,politis2012modeling}. The observational data collected from a variety of sensors can be assumed to comprise of these complex flow interactions and the manifestation of all physical processes. Therefore, the hybrid model will aid in the robustness of the DT rather than a pure physics-based or pure data-driven approach.
}

In the current paper \textcolor{rev}{we view a DT} of a physical asset as the schematic shown in Figure~\ref{fig:DT}. Any physical asset generates data which requires real-time processing for mainly two purposes; informed decision making and real-time optimization and control of the physical asset. Additionally, the recorded data can be perturbed to do offline \emph{what if?} analysis, risk assessment, uncertainty quantification and process optimization. In the latter context it is more appropriate to term the various virtual realizations of the physical asset as \emph{digital siblings}. 

\begin{figure}[ht]
	\centering
	\includegraphics[width=\textwidth]{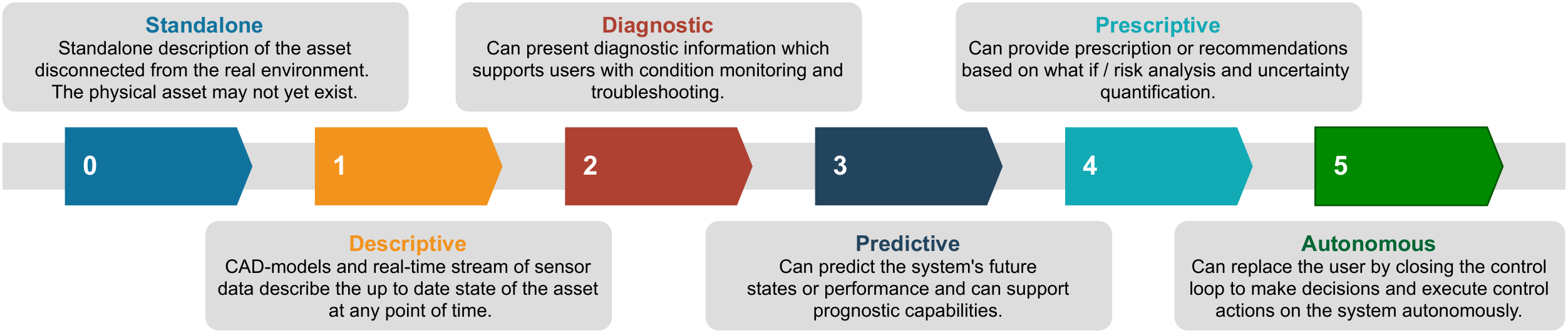}
	\caption{The capability level of DTs on a scale from 0 to 5. We envision that the HAM can leverage simulating tools from 0-2 scales to 3-5 scales.} 
	\label{fig:scales}
\end{figure}

While DTs operating as digital siblings can be used for hypothetical scenarios, \emph{digital threads} can be used for transferring experience from one asset to the next generation assets. As depicted in Figure~\ref{fig:scales}, the capability of DT\cite{dnvgl} can be ranked on a scale from 0 to 5 (0-standalone, 1-descriptive, 2-diagnostic, 3-predictive, 4-prescriptive, 5-autonomy). Once multifidelity modeling tools relevant for specific applications are developed, a DT framework can be utilized for holistic optimization throughout the design, installation, operation, and decommissioning phases. Therefore, the DT representation can facilitate rapid knowledge exchange between \emph{research and innovation} with \emph{design and operations}.       

%With the recent wave of digitalization, the latest trend in every industry is to build systems and approaches that will help it not only during the conceptualization, prototyping, testing and design optimization phase but also during the operation phase with the ultimate aim to use them throughout the whole product life cycle and perhaps much further beyond that. While in the first phase, the importance of numerical simulation tools and lab-scale experiments is not deniable, in the operational phase, the potential of real-time availability of data is opening up new avenues for monitoring and improving operations throughout the life cycle of a product. In our recent survey paper\cite{rasheed2020digital}, we have presented a DT framework and detailed its values, challenges and enabling technologies.  

%The necessity to formalize and utilize the full potential of DT concepts arises from a combination of technology push and market pull. While the need for online monitoring, flexibility in operation, better inventory management and personalisation of services are the most obvious market pull, availability of cheap sensors and communication technologies, phenomenal success of ML and artificial intelligence, in particular, deep learning, new developments in the computational hardware, cloud and edge computing are certainly the major technology push. In this regard, we foresee that the HAM concept is going to bring revolution across an array of applications in numerous industry sectors. 

The communication between the physical asset and its DT can allow for a rapid testing of industry-standard workflows, as well as the state of the art data assimilation and optimal control theories. The chief idea is to establish a two-way coupling in real-time between the physical asset and the simulator through data assimilation and model updates. Experimental measurements can be immediately forwarded and seamlessly integrated into the simulator, while at the same time, the simulator can directly control the physical asset or provide metrics for a better informed decision-making process. Moreover, an experimental facility or set-up can provide a unique possibility for designing, calibrating and validating a DT simulator. Of specific interest, the traditional industrial workflow in many sectors use simulators as a planning and decision-making tool, wherein data are integrated via history matching to improve the simulator's parameterizations. This standard workflow can be significantly improved by the concept defined by Bolton et al. \cite{bolton2018customer} as ``a dynamic virtual representation of a physical object or system across its lifecycle, using real-time data to enable understanding, learning, and reasoning'' that might yield ultimately a software package to allow for real-time control of operating conditions. In practice, the following aspects are of great interest:

\medskip
\emph{Validating the physical model and data integration technologies.}  Due to the availability of precise characterization in complex systems, the skill of data assimilation technologies can be carefully assessed in comparative studies with true physical benchmarks. Moreover, no standard for building a DT exists. Most of the state of the art DT technologies operate on a capability scale 0-2 having little physical realism owing to lack of data and real-time models with the ability to model unknowns. This might allow for a critical review of industry best-practices, as well as evaluation and development of new methodologies.

\medskip
\emph{The potential for automatic control.}  The time scales of possible experimentation (hours to days to weeks) and the ability to repeat these experiments with nearly identical physical features allow for investigation for more improved model-based automatic control strategies. Control algorithms are often linear, use overly simplified models, and rarely optimize for multiple objectives simultaneously. 

\medskip
\emph{Validation of scenario and risk assessment.} Carefully designed scenario and risk assessment studies should be performed with the same physical conditions. A DT framework can be also used to assess sustainability.

\medskip
\emph{Enhanced training.} In many systems, the decisions are often based on intuition obtained from indirect observations or simulations with incomplete parameterizations. The transparency of the front panel may provide direct feedback for the users to correlate between indirect measurements and underlying system dynamics.   

\medskip
\emph{Predictive maintenance.} The component level condition monitoring can be speeded up by using the HAM approaches and then can be integrated together to develop asset level condition monitoring methodologies. Predictive maintenance decisions and remaining lifetime predictions based on condition monitoring and statistical reliability of data interfaced with predictions of broader interconnected systems may set important considerations toward a holistic and probabilistic framework for decision support.      

\medskip
\textcolor{rev}{
\emph{Asset management.} Since DTs often combine different assets and their information in a central place, it will be crucial to ensure that data is visually represented well, secure and available to all relevant stakeholders.
}

\medskip
\emph{Public engagement.} Virtual reality or augmented reality technologies can be utilized to inform the public for the future impacts of the upcoming or hypothetical scenarios for an asset. For example, wind energy developments may face public resistance \cite{barnett2012imagined}. More informed communication of the impact of upcoming wind farms to the public can benefit a lot from DT technologies.

\section{Concluding Remarks and Outlook}\label{sec:conc}
In this perspective paper, we primarily focus on the combined use of physics-based and data-driven models, denoted as HAM, as a science and engineering disruptor that can be exploited to achieve the goals of the development of better predictive methods and tools to form a digital transformation in domain-specific fields. \textcolor{rev}{Figure~\ref{fig:ham} shows a non-inclusive list of various directions where HAM can be pursued.}  The outlined HAM frameworks aim squarely at transforming both physics-based and data-driven methodologies into general, robust and portable computational tools for applications across the spectrum in engineering, science, and medicine.
%Development of HAM algorithms and architectures, which are a core enabler of emerging DT applications, can provide a basis to generate predictive technologies for a broad spectrum of engineering and science applications including pattern classification and scale bridging of hierarchical climate simulations. 
From our point of view, there is an urgent need to focus on HAM algorithms and architectures, which are a core enabler of emerging DT applications. To this end, we identify a set of topics for further exploration below. 

\medskip
\emph{Knowledge-informed ML.} The problem of building a model from the high-fidelity data can be posed as a multi-objective optimization problem where several aspects of the knowledge-injection processes must be involved in data preparation, training and testing steps. When it comes to developing ML algorithms and architectures, this HAM approach is of critical importance to tackle challenges such as incorporating physical laws within the learning framework, producing solutions that are interpretable and explainable, addressing nonlinearities and conservation properties, and dealing with the massive amount of data needed for training. It is possible to gain huge performance increases (e.g., improved generalizability with reducing neural network model uncertainty) when we augment the knowledge of the domain-specific simplified theories with the underlying learning process.

\medskip
\emph{Data preprocessing, compression and superresolution:} Main issues associated with data are noise, missing data, sparse data, and excessively dense data. Although there are state-of-the-art methods for dealing with these issues, a potential research focus in this domain can revolve around generative adversarial networks, and principal component analysis (or its nonlinear extensions \cite{kramer1991nonlinear,hsieh2001nonlinear,monahan2000nonlinear,dong1996nonlinear,mori2016nonlinear}, deep autoencoders \cite{charte2018practical,lee2020model,agostini2020exploration,xu2020multi,otto2019linearly} etc). While the former is a powerful method for synthesizing new data from sparse data, the latter is a dimensionality reduction algorithm useful in discarding irrelevant features in data. 

\medskip
\textcolor{rev}{
\emph{Improving reduced order models:} 
ROMs have great promise for flow control, data assimilation, parameter estimation, and uncertainty quantification. However, their potential has only been realized for a limited set of problems. For example, fluid flows typically involve nonlinear interactions over very wide ranges of length and time scales. Although high-fidelity simulations of complex flows are possible to a certain extent with today’s supercomputers, computationally feasible yet accurate ROMs that operate under such strong nonlinear interactions with a wide range of geometric and physical parameter changes currently do not exist. When it comes to utilizing machine learning algorithms in fluid dynamics, some of the challenges include incorporating physical laws within the learning framework, producing solutions that are interpretable, addressing nonlinearities, and dealing with the massive amount of data needed for training. Therefore, synthesizing ideas from physics-based modeling, numerical methods, data-driven tools, and machine learning can be considered one of the most active areas in ROM community to tackle fundamental barriers not addressed by current, evolutionary modeling practices. These include the curse of dimensionality, lack of accuracy, modal deformation due to transport processes, automatic feature selection, dynamically adapted error metrics, sparse and anisotropic data, sensitivity to noise, and lack of conservation.
}

\medskip
\emph{Discovering new physics using ML and multivariate data analysis:} It is discovered that convolutional layers closer to input modes capture small scale features while the layers closer to the output modes captures large scale features. Neural style transfer is a consequence of this discovery. In fluid mechanics, a potential application of this discovery can lead to a better understanding of turbulence dominated by multiple spatio-temporal scales. Moreover, we can learn constitutive relations from indirect observations using deep neural networks (for example see \cite{huang2020learning}). Alternatively, to address the key limitation of the black-box learning methods, we can exploit to use of symbolic regression as a principle for identifying relations and operators that are related to the underlying system dynamics. This approach combines evolutionary computation with feature engineering to provide us a tool to create new models (e.g., see our recent application for the closure discovery in turbulence \cite{vaddireddy2020feature}). 
The most prohibitive obstacle to overcome in order to search for an appropriate closure or correction model is finding meaningful and nontrivial invariants (e.g., invariants of deformation gradient, strain-rate and vorticity gradient tensors, Lie symmetries, and gauge transformations). 

\medskip
\emph{Using physics and domain knowledge to better frame ML solutions:} How should we inject physics and domain knowledge into AI models? Our recent work in this direction pertains to introducing a physics guided ML approach based on augmenting reason-based and principled models into the neural network architectures. To emphasize on their physical importance, our architecture consists of adding certain features at intermediate layers rather than in the input layer. 
Such a feature enhancement approach can significantly improve the prediction accuracy with better uncertainty confidence characteristics, and it can be effectively used in many scientific ML applications, especially for the systems where we can use a theoretical, empirical, or simplified model to guide the learning module.  In certain applications, like offshore wind energy predictions, we can create a huge benefit of improving the prediction system if the domain knowledge is utilized for feature selection and feature engineering before using a data-driven approach. 

\medskip
\emph{Improving interpretability and explainability of ML algorithms:} Although the universal approximation theorem states that a deep neural network can approximate any complex function, it is imperative to make the neural networks more interpretable for a risk free adoption of the approaches in safety critical applications. A first step to interpreting these networks is to develop alternative representations that allow for further analysis. We have provided a proof-of-concept as a first step for computing piecewise affine form of any fully connected networks. We plan to extend our analysis to modern networks with more complex branching architectures. Using our dissecting method together with dimensionality reduction techniques is an approach that shows great promise for the study of complex systems that resist analysis.     

\medskip
\emph{Communication cost reducing algorithms.} Probably, another important aspect is the design of communication-cost reducing algorithms toward exascale computing with emerging heterogeneous architectures. Since the chip manufacturing sector reaches the limits of the atomic scale, unless we see a new transistor technology to replace current metal-oxide semiconductor technologies, a more effective design of algorithms becomes crucial as the number of processing elements increases and/or the number of grid points decreases within a processing element. This leads to more effective use of transistors through more efficient architectures in an ecosystem of extreme heterogeneity. The HPC community has moved forward to incorporate GPU based accelerators and beyond (e.g., TPUs) for not only graphics rendering but also scientific computing applications in a post Moore's Law world. This heterogeneity shift will become even more crucial in the future since there is a rapid increase in the usage of high-productivity programming languages (e.g., MATLAB, R, Python, Julia) among engineers and scientists. Beyond the obvious use of GPUs in tensor based methods, the integration of traditional HPC practices with ML becomes crucial to obtain major performance improvements. 

\medskip
\emph{Integrating ML into large scale computations:} Where is it valuable and how should we do this? Combining ML methodologies with solvers, a chief research focus can be tackling a series of open cyberinfrastructure questions and challenges that the HAM paradigm presents. We advocate further to exploit techniques that are more common in ``ML for HPC'' than ``HPC for ML'' and investigate the benefit of this approach. This requires a seamless approach of both horizontal (many tasks) and vertical (hierarchical) parallelisms. For example, sampling techniques today offer an unbiased way of quantifying uncertainty. Using ML for pattern recognition in combination with such techniques is a promising option for reliable uncertainty quantification for data sets that are spatially limited and have limited resolution. \textcolor{rev}{Moreover, most data-driven approaches involve matrix operations, which can be very efficiently parallelized on affordable graphical processing units and tensor processing units giving several orders of magnitude speedup as required in real-time applications. Although ``ML for quantum computers'' is largely an open area, utilizing quantum computers to potentially enhance conventional machine learning algorithms has been discussed recently in the context of generating high-quality images \cite{rudolph2020generation}.}

\medskip
\emph{Toward asynchrony-tolerant computations:}   
Increasing the scalability of the codes is a direct consequence of relaxing the data dependence, since the requirement of such global synchronization throughout the domain leads to poor scaling performance with increasing system size, especially for more realistic direct numerical simulations. It is thus imperative to develop computational strategies and stable algorithms that tolerate asynchrony between processing elements. To mitigate the effect of asynchrony, one can exploit the feasibility of a ML based proxy-equation paradigm for asynchrony-tolerant computations. This approach might introduce a data-driven delay correction factor by modifying the time scale near processing element boundaries (i.e., learning time index, adding inertia to the system to reduce the delay error and improve accuracy).  

\medskip
Overall, the unifying theme in these research thrusts is to develop a HAM paradigm that combines physics-based modeling with the versatility of data-driven approaches. The HAM concept consists of a wide spectrum from fundamental research, to enabling technology development, to the system integration phase, and thus offers a compelling theme for center-wide projects as well as university-industry partnership calls. \textcolor{rev}{The interactions between domain specific scientists and data scientists have increased exponentially in recent times. Hence we may witness many novel approaches or even paradigm shifts in the near future to tackle some of the fundamental challenges not addressed by evolutionary segregated modeling practices.} We believe that there is an urgent need to focus on online real-time compression and extrapolatory predictive modeling for safety-critical applications in engineering and applied sciences. Through the development of such a profound HAM computational pipeline, we believe that it is possible to advance predictive tools for simulating and characterizing complex multi-x systems with quantified uncertainties and make them more trustworthy, generalizable, and explainable for real life applications.

%\backmatter

\section*{Acknowledgments}
%This is acknowledgment text~\cite{Elbaum2002}. Provide text here. This is acknowledgment text. Provide text here. This is acknowledgment text. Provide text here. This is acknowledgment text. Provide text here. Provide text here. This is acknowledgment text. Provide text here. This is acknowledgment text. Provide text here. 
%Omer San's work has been supported by the U.S. Department of Energy, Office of Science, Office of Advanced Scientific Computing Research under Award Number DE-SC0019290 and the National Science Foundation under Award Number DMS-2012255. He gratefully acknowledges their support. Adil Rasheed and Trond Kvamsdal gratefully acknowledge the funding received by SINTEF Digital in the SEP Digital Twin project (102015968-22). 
%The authors would like to thank XXX for for his insights on this manuscript.
Omer San would like to acknowledge support from the U.S. Department of Energy under the Advanced Scientific Computing Research program (grant DE-SC0019290), the National Science Foundation under the Computational Mathematics program (grant DMS-2012255). Adil Rasheed and Trond Kvamsdal gratefully acknowledge the funding received by SINTEF Digital in the SEP Digital Twin project (102015968-22). OPWIND: Operational Control for Wind Power Plants (Grant No.: 268044/E20) project funded by the Norwegian Research Council and its industrial partners (Equinor, Vestas, Vattenfall) is also acknowledged.

%\subsection*{Author contributions}
%The authors contributed equally to this work.

%This is an author contribution text. This is an author contribution text. This is an author contribution text. This is an author contribution text. This is an author contribution text. 

%\subsection*{Financial disclosure}

%None reported.
%\subsection*{Compliance with ethical standards}
%All procedures followed were in accordance with the ethical standards of the responsible committee on human experimentation (institutional and national) and with the Helsinki Declaration of 1975, as revised in 2000. Informed consent was obtained from all patients for being included in the study.

\subsection*{Conflict of interest}
The authors declare no potential conflict of interests.

\bibliography{references}

\newpage
\section*{Author Biography}

\begin{biography}{\includegraphics[width=60pt,height=70pt, keepaspectratio]{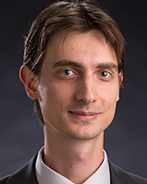}}{\textbf{Omer San} is an assistant professor of Mechanical and Aerospace Engineering at Oklahoma State University, Stillwater, OK, USA. His expertise is in fluid dynamics with a special focus on multiscale modeling, high performance computing, scientific machine learning, and data-driven modeling approaches. 
%His expertise is in computational science with a special focus on high performance computing, scientific machine learning, big data cybernetics and data-driven modeling approaches. 
%His approach is broad, using theoretical, computational, and statistical techniques, with an emphasis on developing a hybrid analysis and modeling  framework, combining physics-based models with the versatility of data-driven approaches.
%His field of study is centered upon the development, analysis and application of advanced computational methods in science and engineering with a particular emphasis on fluid dynamics across a variety of spatial and temporal scales. 
He received his bachelors in Aeronautical Engineering from Istanbul Technical University in 2005, his masters in Aerospace Engineering from Old Dominion University in 2007, and his Ph.D. in Engineering Mechanics from Virginia Tech in 2012. He worked as a postdoc at Virginia Tech from 2012-'14, and then from 2014-'15 at the University of Notre Dame, Indiana. He is a recipient of U.S. Department of Energy 2018 Early Career Research Program Award in Applied Mathematics.}
\end{biography}

\begin{biography}{\includegraphics[width=60pt,height=70pt, keepaspectratio]{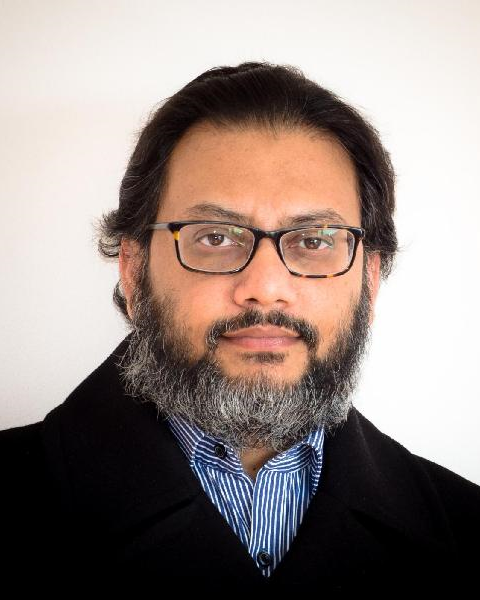}}{\textbf{Adil Rasheed} is a professor of Big Data Cybernetics in the Department of Engineering Cybernetics at the Norwegian University of Science and Technology where he is working to develop novel hybrid methods at the intersection of big data, physics-driven modeling and data-driven modeling in the context of real-time automation and control. He also holds a part time senior scientist position in the Department of Mathematics and Cybernetics at SINTEF Digital where he led the Computational Sciences and Engineering group between 2012-2018. He holds a PhD in Multiscale Modeling of Urban Climate from the Swiss Federal Institute of Technology Lausanne. Prior to that he received his bachelors in Mechanical Engineering and a masters in Thermal and Fluids Engineering from the Indian Institute of Technology Bombay.}
\end{biography}

\begin{biography}{\includegraphics[width=60pt,height=70pt, keepaspectratio]{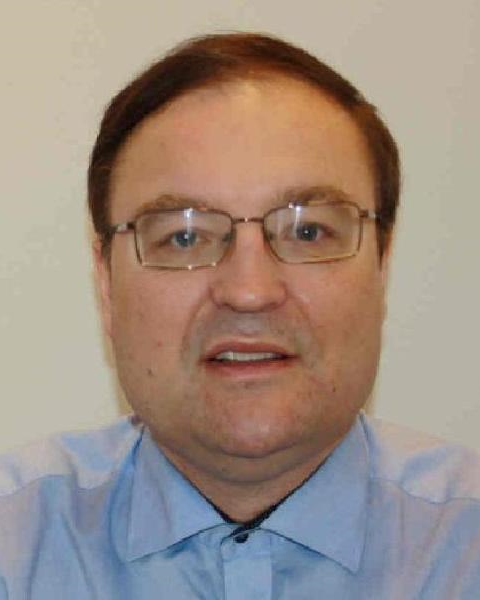}}{\textbf{Trond Kvamsdal} is a professor in the Department of Mathematical Sciences, Faculty of Information Technology and Electrical Engineering at the Norwegian University of Science and Technology. His positions at NTNU are within computational mathematics, i.e. development of new theories/methods within applied mathematics and numerical analysis to make robust and efficient numerical software programs for challenging applications in science and technology. Main area of application is computational mechanics, i.e. both solid/structural and fluid mechanics relevant for civil, mechanical, marine, and petroleum engineering as well as biomechanics, geophysics and renewable energy. He received the IACM Fellow Award (International Association for Computational Mechanics) in 2010 and was elected
member of the Norwegian Academy of Technological Sciences (NTVA) in 2017.}
\end{biography}

% \begin{biography}{\includegraphics[width=60pt,height=70pt,draft]{empty}}{\textbf{Author Name.} This is sample author biography text this is sample author biography text this is sample author biography text this is sample author biography text this is sample author biography text this is sample author biography text this is sample author biography text this is sample author biography text this is sample author biography text this is sample author biography text.}
% \end{biography}

\end{document}